\documentclass[twocolumn]{aastex631}
\usepackage{amsmath}
\usepackage{physics}
\DeclareMathOperator{\sign}{sgn}
\usepackage{appendix}

\shorttitle{Simulating Solar Neighborhood Brown Dwarfs}
\shortauthors{Honaker et al.}

\graphicspath{{./}{figures/}}

\begin{document}

\title{Simulating Solar Neighborhood Brown Dwarfs I:\\ The Luminosity Function Above and Below the Galactic Plane}

\correspondingauthor{Easton Honaker}
\email{ehonaker@udel.edu}

\author[0000-0003-1202-3683]{Easton J. Honaker}
\affiliation{Department of Physics and Astronomy, University of Delaware, Newark, DE 19716, USA}

\author[0000-0002-8916-1972]{John E. Gizis}
\affiliation{Department of Physics and Astronomy, University of Delaware, Newark, DE 19716, USA}

\begin{abstract}
    Brown dwarfs form the key, yet poorly understood, link between stellar and planetary astrophysics. These objects offer unique tests of Galactic structure, but observational limitations have inhibited their large-scale analysis to date. Major upcoming sky surveys will reveal unprecedented numbers of brown dwarfs, among even greater numbers of stellar objects, greatly enhancing the statistical study of brown dwarfs. To extract the comparatively rare brown dwarfs from these massive datasets, we must understand the parameter space they will occupy. In this work, we construct synthetic populations of brown dwarfs in the Solar Neighborhood to explore their evolutionary properties using Gaia-derived star formation histories alongside observational mass, metallicity, and age relationships. We apply the Sonora Bobcat, SM08, and Sonora Diamondback evolutionary models. From the populations, we explore the space densities and median ages by spectral type. We present the simulated luminosity function and its evolution with distance from the Galactic Plane. Our simulation shows that brown dwarf population statistics are a function of height above/below the Galactic Plane and sample different age distributions. Interpreting the local sample requires combining evolutionary models, the initial mass function, the star formation history, and kinematic heating. Our models are a guide to how well height-dependent samples can test these scenarios. Sub-populations of brown dwarfs farther from the Plane are older and occupy a different region of parameter space than younger sub-populations closer to the Galactic Plane. Therefore, fully exploring population statistics both near and far from the Plane is critical to prepare for upcoming surveys.  
\end{abstract}


\section{Introduction}\label{INTRODUCTION}
The study of brown dwarfs, sub-stellar objects that lack the ability to fuse hydrogen, has been driven by large-scale digital sky surveys for the past 30 years. Originally theorized in 1962 \citep{Kumar}, brown dwarfs have been discovered through dedicated surveys with large facilities such as 2MASS \citep{2mass,kirkpatrick99Ltype, gizis20002mass}, PAN-STARRS1 \citep{ps1,deacon14ps1,best18ps1}, WISE \citep{wise,Cushing11Y,Kirkpatrick2011Wise}, DENIS \citep{denis,Delfosse_Denis}, SDSS \citep{sdss,west04}, and UKIRT \citep{ukirt,burningham10}, but also as foreground contaminants in extragalactic surveys \citep{Ceers,tee_predicting_2023}. We are entering an exciting time for brown dwarf science as the next generation of observation facilities become a reality and start to come online. Observatories such as the James Webb Space Telescope (JWST), Euclid, the Vera C. Rubin Observatory, and the Nancy Grace Roman Space Telescope promise to drastically increase the discovery rate for brown dwarfs and propel the field forward. 

Brown dwarfs are cold, sub-stellar objects that extend beyond the main sequence through spectral types L \citep{martin99L, kirkpatrick99Ltype}, T \citep{burgasser02T,geballe02T}, and Y \citep{delorme2008Y,Cushing11Y}. These objects have effective temperatures below $\sim$2400, 1300, and 600 K, respectively. As such, these objects are faint in wavelengths shorter than 1 $\mu$m and present a challenge to observe. Furthermore, since brown dwarfs are not massive enough to fuse hydrogen in their cores, they perpetually cool and grow dimmer, only increasing the difficulty of their observation.

Brown dwarfs are excellent probes for tracing Galactic evolution and structure \citep{burgassergalactic}. Their kinematics have been used to study their population distributions as well as mapping out the structure of the Milky Way’s thin disk \citep{Faherty_BDKP1, kirkpatrick24}. Furthermore, distant M, L, and T dwarfs can help measure thin disk structure and constrain the thick disk and halo scale heights \citep{dupuy17, Aganze_scales, Best_2024}. Therefore, the ability to detect more distant brown dwarfs and understand their formation mechanisms with current and future surveys will play a key role in mapping the Milky Way and advancing such science cases.

Accurately modeling the atmospheres of brown dwarfs is a complicated affair. While their interiors are well known to be fully convective, their atmospheres are complex, filled with molecular features and weather that depend on temperature and pressure evolution \citep{burrows06}. The warmer L dwarfs are covered by cloudy atmospheres, but the observation of these clouds is inclination-dependent \citep{suarez23inclination}. As brown dwarfs cool through the L sequence and into the T sequence, the clouds sediment and ``rain out'' minerals and metals leading to clear, cloudless atmospheres \citep{burrows06}. These T dwarfs are further characterized by strong molecular features and the emergence of methane bands in their spectra. Finally, the coolest Y dwarfs spectral energy peaks at $\sim$5 $\mu$m and show large spectral absorption features from molecules like water, methane, ammonia, carbon monoxide, and carbon dioxide \citep{Cushing11Y}. It appears increasingly likely that late-T and Y dwarf spectra must be described by a second parameter other than temperature, such as metallicity or surface gravity \citep{beiler23JWSTy}. These diverse spectral types are not distinct classifications without overlap, but rather form an evolutionary sequence brown dwarfs undergo as they age and cool. The cooling mechanisms of brown dwarfs are an area of active research for future surveys. The resulting luminosity function is a key observational constraint on the evolutionary models describing such cooling mechanisms \citep{best21, Kirkpatrick_2021_IMF, Best_2024, kirkpatrick24}.

In order to prepare for next-generation surveys that will reveal these dim populations of brown dwarfs and enable the statistical analysis of their characteristics, we must understand their underlying distribution within the Galaxy. We turn to simulating the Solar Neighborhood population of brown dwarfs to predict the characteristics of brown dwarfs that future surveys will see. Previously, the creation of synthetic populations required assuming a star formation rate, initial mass function (IMF), and an underlying exponential distribution with height above the Galactic Plane \citep{ryan_self-consistent_2022}. We present a novel approach using recent Gaia results in conjunction with recent observational relations to build synthetic populations of brown dwarfs.

This paper presents our simulation of Solar Neighborhood brown dwarfs. In Section 2 we build the synthetic population of brown dwarfs. We present the results of the simulation as parameter functions in Section 3. The implications of the results are discussed and compared to recent studies in Section 4. Our conclusions and future work are summarized in Section 5. 

\section{Simulations}\label{SIMULATIONS}
We simulate synthetic populations of brown dwarfs in the Solar Neighborhood using three different evolutionary models. Starting from observational relations, we assemble the population and apply the evolutionary models to determine the variation of parameters with height above and below the Galactic Plane. Each step in the process is described in detail in the following subsections.

We note that, historically, the Solar Neighborhood refers to a spherical volume radially extending tens to hundreds of parsecs from the Sun \citep{Henry_SN_94,GaiaNSC}. In this work, we simulate a cylindrical volume that extends above and below the Galactic Plane, but we refer to it as the Solar Neighborhood for simplicity and to emphasize the importance of the growing resolvable volume with future surveys.

\subsection{Base Population}
\label{basecatalog}
Recent analysis of the Solar Neighborhood using Gaia by \citet{mazzi_SFR_2024} has revealed the star formation rate (SFR) history for a cylindrical volume centered around the Sun.  The cylinder has a radius of 200 pc, is centered on the Galactic Plane, and extends $\sim$1300 pc above and below the Plane. This SFR represents the star formation history that produced stellar populations as they are observed today for a given height above and below the Plane. As a result, this does not represent where the stars originally formed, but accounts for kinematic effects. We assume a universal initial mass function such that the brown dwarf formation history follows the stellar formation history.

\begin{figure}
    \centering
    \includegraphics[width=\columnwidth]{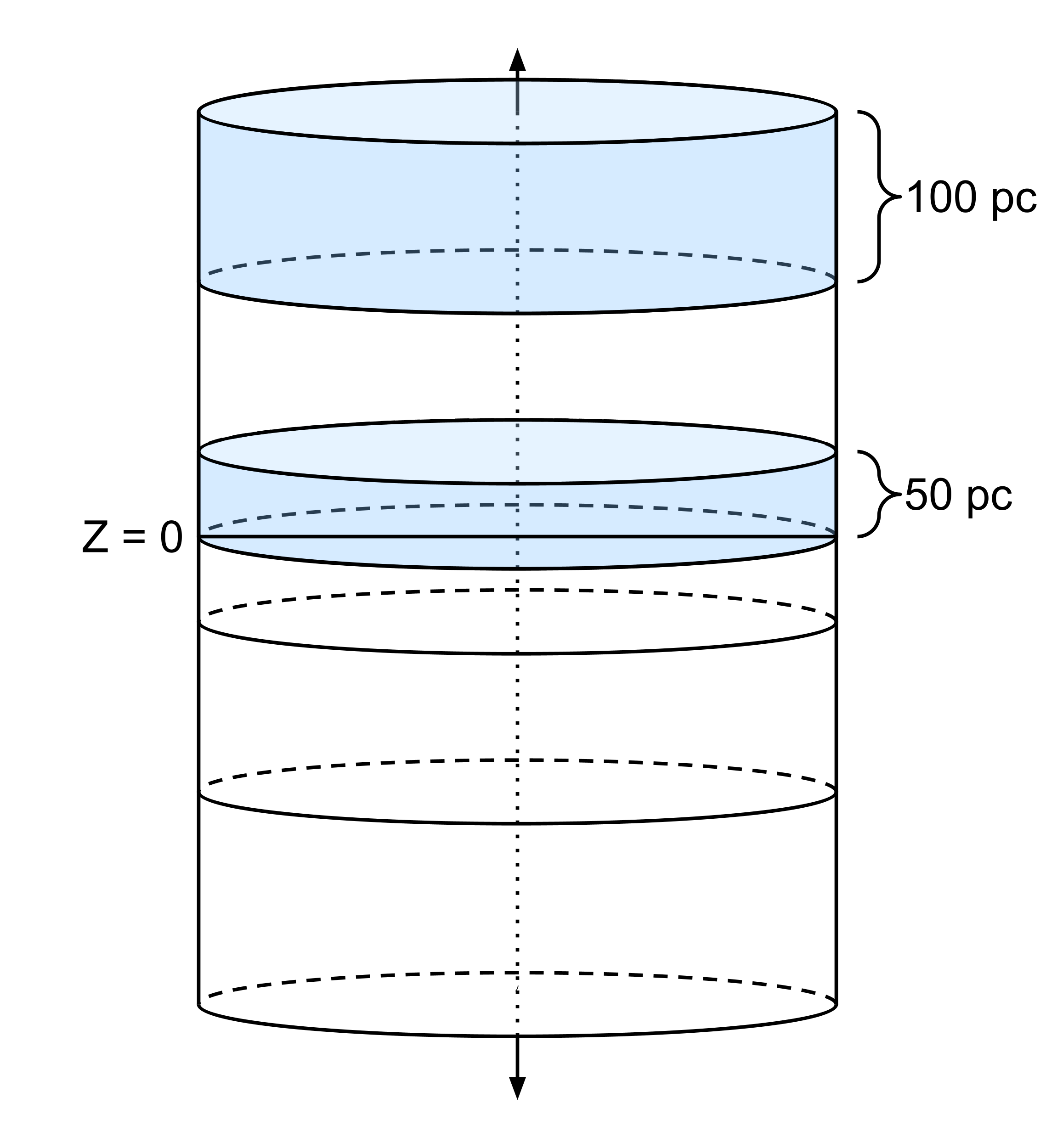}
    \caption{The simulated Solar Neighborhood cylinder with a radius of 200 pc and total height of 2600 pc, 1300 pc above and below the Galactic Plane ($z=0$). The $z$-axis is aligned with the cylinder's vertical axis of symmetry. We divide the cylinder into horizontal slices with smaller $\Delta z = 50$ pc close to the Galactic Plane and more distance slices with larger increments $\Delta z = 100$ pc. This difference allows for higher spatial resolution near the Galactic Plane.}
    \label{cylinderdiagram}
\end{figure}

Positions within the cylinder are denoted by $(x,y,z)$ where $x$ points toward the Galactic center, $y$ points along the Milky Way rotation, and $z$ is perpendicular to the Galactic Plane. We adopt the same coordinate system with the Galactic Plane at $z = 0$ pc and the Sun at $(x,y,z) = (0,0,17.7)$ pc, consistent with previous studies (\cite{mazzi_SFR_2024}, and references therein). The reported SFR history is given as a function of age and $z$. Temporal bins range from $\log$(t [yr]) $\in [6.6,10.10]$ with bin widths of $\Delta \log(t) = $ 0.2 for all bins except the first bin, which extends from 6.6 to 7.1.  The cylinder is divided into horizontal discs by height above/below the Plane, as shown in Figure \ref{cylinderdiagram}. Physical slices are in intervals of $\Delta z \approx 50$ pc for $\abs{z} \lesssim 150$ pc and $\Delta z \approx 100$ pc for $150 $ pc $ \lesssim \abs{z} \lesssim 1300$ pc. In total, there are 16 temporal slices and 28 spatial slices (14 on each side of the Galactic Plane). 

For each time and space bin, \citet{mazzi_SFR_2024} report a SFR in $M_\odot$ yr$^{-1}$ kpc$^{-3}$. We normalize the total star formation for each spatial slice (total SFR across all age bins for a given physical slice) using the space density of $1.83 \times 10^{-2}$ pc$^{-3}$ from \citet{Best_2024}, derived from the complete sample of L0 and later dwarfs within 25 pc. We apply this space density to the Solar slice ($0 \leq z \leq 52.63$ pc) and normalize the total SFR for all other slices such that the Solar slice matches observed total space densities. 

\begin{figure}
    \centering
    \includegraphics[width=\columnwidth]{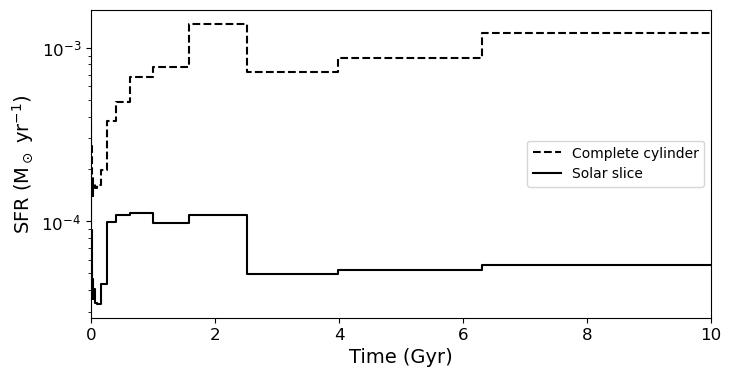}
    \caption{The volume-integrated star formation rate for the Solar slice and complete cylinder. Beyond 2 Gyr ago, the SFR in both cases is nearly identical, but more recently the SFR shows significant differences, with bursty star formation in the solar slice. The differences in the spatially-correlated SFR underscore the need to use the Gaia-based SFR as opposed to a uniform SFR across the entire cylinder.}
    \label{sfr}
\end{figure}

In Figure \ref{sfr}, we show the volume-integrated SFR for the Solar slice and for the entire cylinder. While the full cylinder and the Solar slice have analogous SFR from 3 - 10 Gyr, the Solar slice has a clearly different recent history with a higher proportion of young objects. The spatially-correlated differences in recent star formation demonstrate the importance of using the Gaia-derived SFR history instead of assuming a uniform star formation rate.

The total number of simulated objects per physical slice is determined by the normalized SFR. Positions are randomly drawn from a uniform distribution within a physical slice and object ages are calculated by proportionally dividing the total number of objects within the physical slice by the SFR for each temporal bin. Within a temporal bin, object ages are uniformly distributed.

\begin{figure}
    \centering
    \includegraphics[width = \columnwidth]{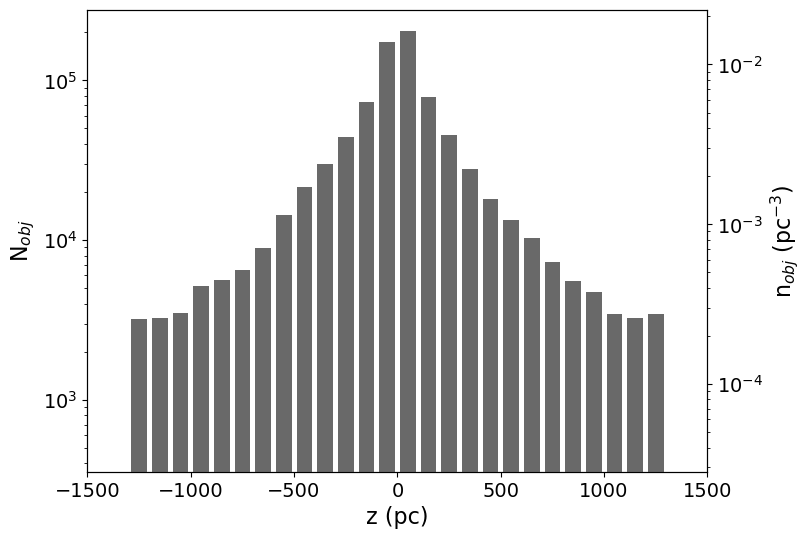}
    \caption{The number and spatial densities of simulated objects per 100 pc interval are shown. The number of objects per interval was obtained by normalizing the SFR history to the local sample. Objects are concentrated at low $\abs{z}$ values and have lower spatial densities farther from the Plane.}
    \label{objectsbyslice}
\end{figure}

Object positions in $(x,y,z)$ are translated to Galactic $l, b$ and distance from the Sun (D$_{obs}$) where:
\begin{equation}\label{Dlb}
    \begin{split}
        D_{obs} = \sqrt{x^2 + y^2 + (z - 17.7)^2},\\
        b = \arcsin{((z - 17.7) / D_{obs})},\\
        l = \sign{(y)} * \arccos{(x / \sqrt{x^2 + y^2})}.
    \end{split}
\end{equation}
In cases where $(x,y) = (0,0)$, $l$ is undefined and we manually set $l = 0$\textdegree. Further transformations from Galactic to other coordinate frames, such as the standard RA/Dec (ICRS), are done through \texttt{Astropy}\footnote{\url{https://www.astropy.org/}} \citep{astropy2013, astropy2018, astropy2022}. The distribution of objects in 100 pc bins is shown in Figure \ref{objectsbyslice}, where slices closest to the Galactic Plane have a higher number density as set by the normalized SFR.

The final two base parameters of the synthetic population are object masses and metallicities. Masses are distributed based on the Initial Mass Function (IMF) from \cite{Kirkpatrick_2021_IMF}:
\begin{equation}\label{masspowerlaw}
    \frac{dN}{dM} \propto M^{-\alpha},
\end{equation}
where $\alpha = 0.6 \pm 0.1$. We allow masses to range from $0.003 - 0.08 $ M$_\odot$.

Metallicities are assigned to the population using the Age-Metallicity Relation (AMR) from \cite{daltio_AMR_2021}: 
\begin{equation}\label{AMR}
    [Fe/H] = \beta(t - 4.5 \mbox{ Gyr}),
\end{equation}
where $\beta = -0.4$ dex$/12$ Gyr. We include Gaussian scatter with $\sigma=0.1$. This AMR assigns solar metallicities to solar-aged objects whereas older objects are more metal-poor.

The SFR, IMF, and AMR can be combined to obtain object positions, ages, masses, and metallicities, the base parameters of our synthetic populations. The total number of objects in the base simulation, normalized using the local brown dwarf volume density, is $\sim$750,000. The objects form a complete simulated cylinder around the Solar Neighborhood, but lack evolutionary parameters such as effective temperatures, radii, surface gravities, and luminosities. 

\subsection{Evolutionary Model Application}\label{applymodels}

From the base catalog created in Section \ref{basecatalog}, we apply three different models to ``evolve'' the synthetic population\footnote{To avoid simulating objects outside spectral types L0 - Y2, and therefore outside the space densities from \citet{Best_2024} used for normalization, we oversimulate the sample before applying evolutionary models and randomly select the appropriate number of objects per slice with spectral types between L0 - Y2, per the temperature-spectral type relation in \citet{Kirkpatrick_2021_IMF}.}. We apply two of the Sonora substellar atmosphere models: Bobcat \citep{SonoraBobcat_2021} and Diamondback \citep{SonoraDiamondback_2024}. The third model we apply is the 2008 hybrid model from \citet{SM08} (hereafter SM08). The Sonora Diamondback and Bobcat models are both available through Zenodo (Diamondback: \citet{diamondbackzenodo}; Bobcat: \citet{bobcatzenodo}), and we obtained the SM08 model through correspondence with the authors.

The Bobcat models describe cloudless, substellar objects with near-solar metallicities. The Diamondback models are a hybrid model that includes a gravity-dependent transition from cloudy to cloudless models between 1300 and 1000 K. Below 900 K, the Diamondback evolutionary models use the Bobcat models. The Sonora atmospheric model suite provides atmospheric structure, spectra, chemistry, and evolutionary tables for substellar objects. The evolutionary tables contain effective temperature, radius, luminosity, surface gravity, mass, age, and metallicity.  The SM08 models are similar, incorporating a hybrid transition from cloudy to cloudless from 1400 to 1200 K, but the SM08 evolutionary model assumes solar metallicity and does not have a model spectral grid. 

For each evolutionary model, we linearly interpolate the ages, masses, and metallicities of the base population to retrieve effective temperatures, radii, surface gravities, and luminosities. For the SM08 evolved population, we only interpolate the base population's ages and masses since solar metallicity is assumed.

 From the three evolved populations, we explore how different parameters, such as luminosity and age, vary as a function of $z$.

\section{Results}\label{RESULTS}
In this section, we present the outputs of our simulated Solar Neighborhood population. We explore the distribution of objects and their parameters for all three synthetic populations. The variation of evolutionary parameters with height above and below the Galactic Plane, $\abs{z}$, is of particular interest.

\subsection{Age Distribution}
\begin{figure}
    \centering
    \includegraphics[width = \columnwidth]{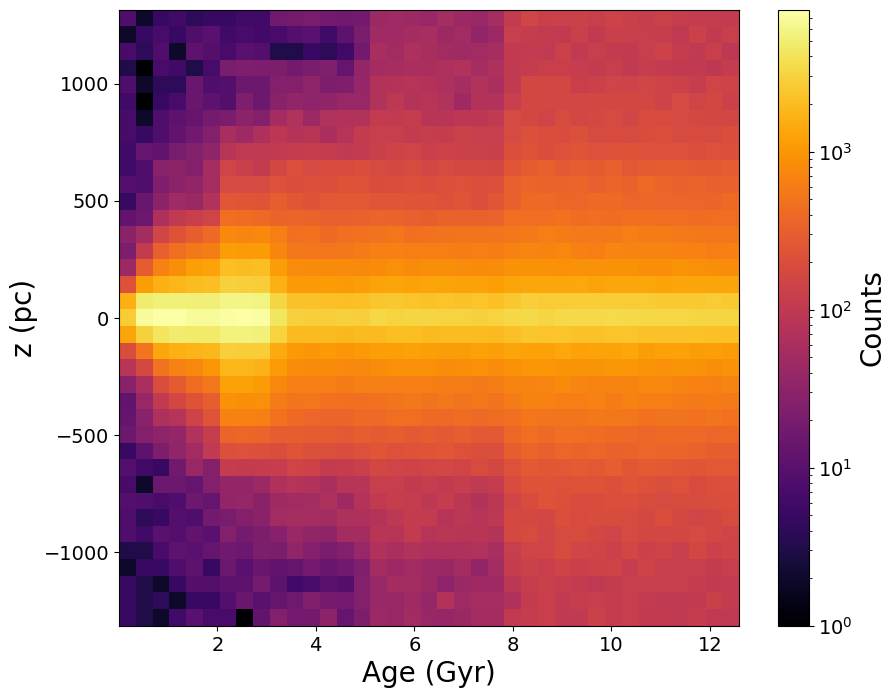}
    \caption{The age distribution of the base population as a function of distance from the Galactic Plane, $z$, is shown. The same age distribution is used for all three synthetic populations. Ages were assigned directly from the SFR history; for a given age, $z$ positions with higher counts are a direct result of higher SFR.
    The age distribution shows a decline in SFR between 4 - 8 Gyr ago with a recent burst of star formation around the Galactic Plane.}
    \label{age_heatmap}
\end{figure}

Object ages form part of the base parameters for the simulation and are identical for all three synthetic populations. In general, objects with smaller $\abs{z}$ values, i.e. closer to the Plane, are younger whereas more distant objects are older. This trend is true for all time bins as seen in Figure \ref{age_heatmap}. This age distribution stems directly from the underlying SFR. The SFR shows a decrease from 4 to 8 Gyr ago before undergoing a recent burst of star formation from 3 Gyr to the present, which is reflected in the age distribution. Figure \ref{age_heatmap} shows older objects ($\geq$ 7 Gyr) are more evenly distributed across $z$ values whereas younger objects ($\leq$ 3 Gyr) are almost exclusively found within 300 pc of the Galactic Plane. The distribution of young and old objects supports an interpretation of dynamical heating of the Galactic Plane, which removes objects from the Plane over time, spreading older populations out to higher $\abs{z}$ values \citep{spitzer53,lacey84,sellwood02, ma17, dupuy17, Best_2024}. 

\subsection{Temperature Function}

\begin{figure}
    \centering
    \includegraphics[width=\columnwidth]{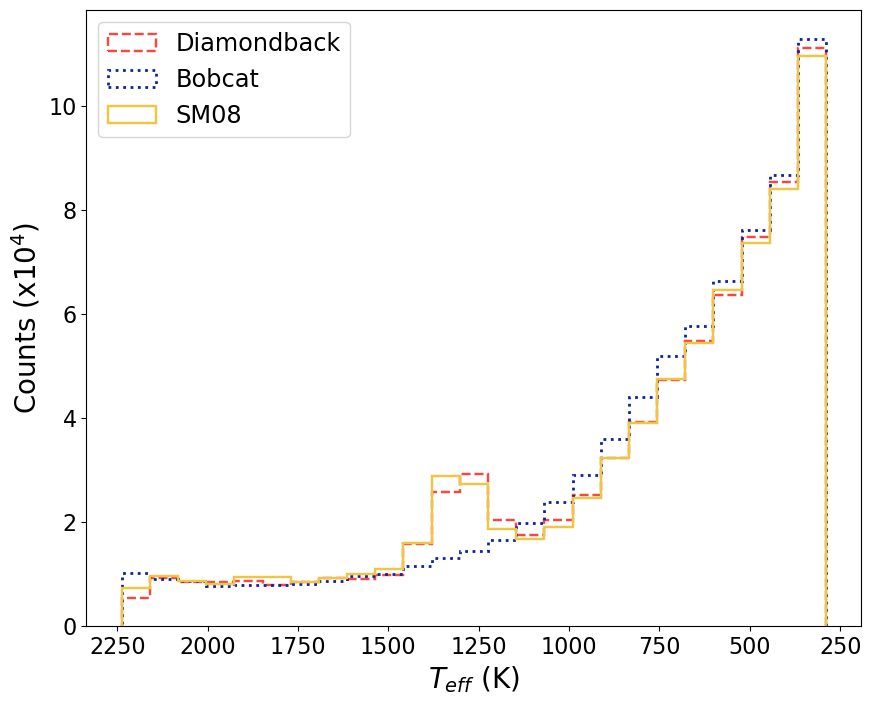}
    \caption{Effective temperature distribution for all three synthetic populations over the entire simulated volume. The hybrid Diamondback (dashed red) and SM08 (solid gold) models show an object pileup as the models transition from cloudy to clear atmospheres. The central peak and width of these features differ based on the interpolation-based and gravity-dependent approaches taken by SM08 and Diamondback, respectively. All three models favor cooler objects below 1000 K.}
    \label{cumulativeTeffdist}
\end{figure}

Object temperatures are obtained by linearly interpolating the base population's age, mass, and metallicity for both Sonora models and interpolating the base population's age and mass for the SM08 model. The resulting distributions of all simulated object temperatures are shown in Figure \ref{cumulativeTeffdist}.

All three synthetic populations temperature distributions favor lower temperatures; this is expected since brown dwarfs perpetually cool and the IMF favors lower mass objects, which in turn leads to colder objects through the evolutionary model tracks. Interpolated Sonora Bobcat population temperatures are smoothly distributed, favoring lower temperatures. The SM08 derived temperatures also favor lower temperatures but have an additional feature centered at 1400 K, surrounding the transition point between cloudy and cloudless atmospheres. This pileup was predicted and discussed in detail by \cite{SM08}. Briefly, the transition pileup occurs as the evolution from L dwarfs to T dwarfs slows down while the L dwarfs adiabatically cool with clouds; an object with clouds requires a longer period of time to release the same amount of energy as an object without clouds. The pileup is nonexistent in the Sonora Bobcat temperatures as the models are exclusively cloudless. The Sonora Diamondback models show a prominent transition feature more sharply concentrated at 1300 K. The Diamondback cloudy-clear atmospheric transition occurs at a cooler temperature than the SM08 transition. 

At temperatures cooler than the atmospheric transition pileups, the Bobcat and Diamondback models are similar, with the Bobcat model predicting slightly more of the coldest objects (T$_{\mbox{eff}} < 400$ K) than the Diamondback model. We only consider objects through spectral type Y2 ($\sim$ 290 K). The SM08 model grid does not extend below T$_{\mbox{eff}} = 275$ K. As such, linear interpolation for the coldest objects results in an unphysical pileup at the bottom edge of the model grid. While this numerical artifact does not appear in Figure \ref{cumulativeTeffdist}, it does appear at the low-luminosity end of Figure \ref{sm08_lum_heatmap} and should be interpreted as a numerical artifact, not a physical effect. The two Sonora models both extend to 200 K and do not encounter this issue.

\subsection{Luminosity Function}
The luminosity function is arguably the most important of the parameter functions as it is the only one that is directly observable; in practice, ultracool dwarf parameters are obtained by measuring an object's luminosity, estimating or assuming an age, and applying evolutionary models to recover remaining parameters \citep{dupuy17,kirkpatrick24}. Additionally, synthetic luminosity functions can be compared to observational luminosity functions to constrain ultracool dwarf cooling mechanisms \citep{Burgasser07,bardalez19,Best_2024, kirkpatrick24}. 

\begin{figure}
    \centering
    \includegraphics[width=\columnwidth]{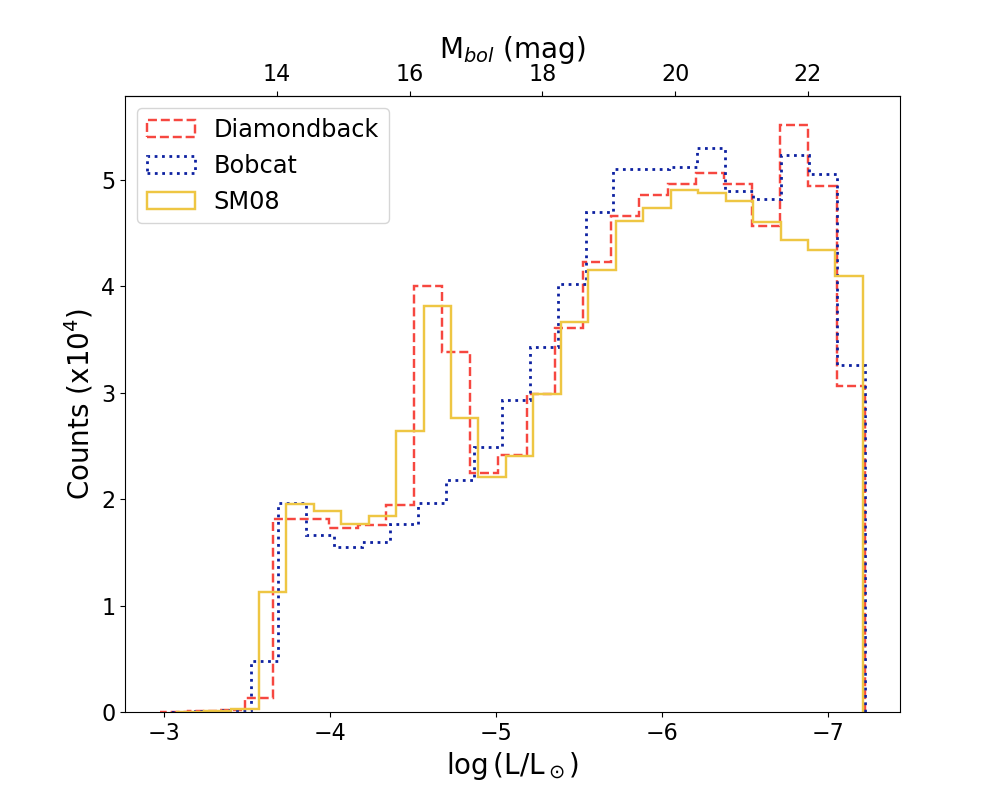}
    \caption{Luminosity and bolometric magnitude distributions for all three synthetic populations over the entire simulated volume. Both the Diamondback (dashed red) and SM08 (solid gold) hybrid models show an atmospheric transition pileup while the cloudless Bobcat models (dotted blue) do not. The Diamondback and Bobcat models show an additional feature at $M_{bol} = 22$ mag unseen in the SM08 models. All three models favor objects with lower luminosities between log(L/L$_\odot) = -5.5$ and $ - 7.25$.}
    \label{cumulativeLumdist}
\end{figure}

In Figure \ref{cumulativeLumdist}, we display the luminosity function of all three synthetic populations for the entire simulated volume. 
To facilitate comparison with previous studies and observations, we also plot the luminosity function in bolometric magnitudes. We converted interpolated luminosities to bolometric magnitudes using M$_{bol,\odot} = 4.740$ mag: 
\begin{equation}
    M_{bol} = M_{bol, \odot} - 2.5 \log(L/L_\odot).
\end{equation}

As shown in Figure \ref{cumulativeLumdist}, both hybrid models have prominent transition pileup features before favoring less luminous objects. The amplitude and width of the pileup feature is significantly larger for the Diamondback model than the SM08 model. These differences reflect the gravity-dependent nature of the Diamondback transition, where the transition phase lasts longer for lower surface gravities. For both the SM08 and Diamondback models, the transition occurs at log(L/L$_\odot$) $\sim$ -4.7 or $M_{bol} \approx 16.25$ mag. All three models favor lower luminosities and have increased counts concentrated around log(L/L$_\odot$) $\approx$ -6.3, or $M_{bol} \approx 20.5$ mag. The SM08 models do not extend below log(L/L$_\odot$) = -7.3, or $M_{bol} = 23$ mag.

Recent analysis of ultracool dwarfs within 25 pc by \citet{Best_2024} yields an observed bolometric luminosity function that is flat from $M_{bol} = 16 - 20$ mag and then peaks from $M_{bol} = 20 -22$ mag. However, magnitudes fainter than $M_{bol} = 21$ mag are not well measured to date. We compare our total volume luminosity function (Figure \ref{cumulativeLumdist}) as well as the luminosity function including only objects within 25 pc of the Sun and in both cases find that our luminosity functions do not reproduce the observed flat region between $M_{bol} = 16 - 20$ mag, but do reach a maximum by $M_{bol} \approx 20$ mag and favor lower luminosities, as suggested by \citet{Best_2024}.

An additional pileup feature in the Diamondback distribution can be seen at $M_{bol} = 22$ that may be related to the transition from T to Y spectral types. This feature is larger than the Poisson noise of the simulation and also appears in the Sonora Bobcat population, albeit smaller in amplitude. From effective temperature-spectral type relations in \citet{Kirkpatrick_2021_IMF}, Y dwarfs have $T_{\mbox{eff}} \leq 460$ K. For the Diamondback model, objects with temperatures below $460$ K have log(L/L$_\odot$) $\leq -6.5$. One possible physical explanation of the secondary pileup feature in the Sonora-based populations is the condensation of water clouds in Y dwarf atmospheres following the T-Y transition. Water condensation below 400 K results in clearer atmospheres as water molecules consolidate into optically thick clouds and no longer act as the dominant atmospheric absorber \citep{Morley14b}. The water condensation allows cooler Y dwarfs to remain more luminous for longer periods of time, leading to a secondary pileup effect below $\sim$400 K similar to, but weaker than, the L-T transition pileup \citep{SonoraBobcat_2021}. However, as stated in \citet{SonoraBobcat_2021}, this interpretation should be used with caution as the Bobcat models do include water condensation effects but not the associated cloud opacities that follows. 

\begin{figure}
    \centering
    \includegraphics[width = \columnwidth]{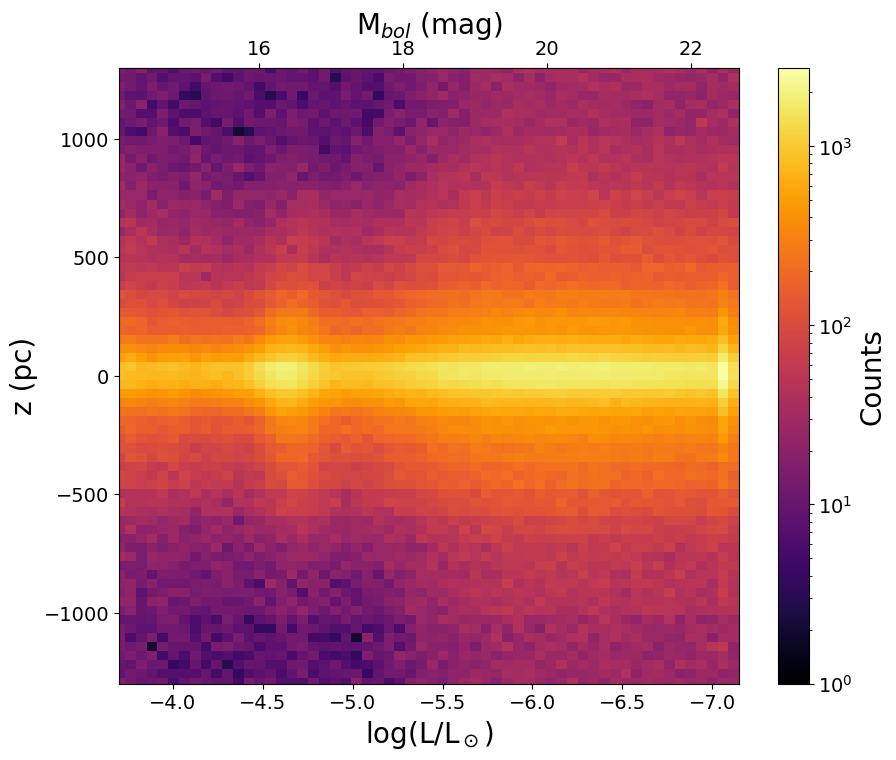}
    \caption{The SM08 synthetic population's luminosity function with respect to position above/below the Galactic Plane. Bolometric magnitudes are also shown for convenience. The pileup from atmospheric modeling transitions is prominently visible at $\log(L/L_\odot) \approx -4.7$ and extends approximately 500 pc both above and below the Plane before gradually disappearing. A smaller concentration of objects is seen at the lowest luminosities ($M_{bol} \geq 22$ mag) near the Plane, however this grouping of objects is likely a non-physical pileup at the edge of the model grid.}
    \label{sm08_lum_heatmap}
\end{figure}

One advantage to our simulation is we are able to show the distribution of luminosities as a function of position above/below the Galactic Plane, as shown in Figure \ref{sm08_lum_heatmap} for the SM08 population (see Figure \ref{luminositywithpopouts} for the Sonora model-based populations). Plotting the luminosity function as a function of $z$ allows us to separate the contributions from different distances above/below the Galactic Plane.  The same features seen in Figure \ref{cumulativeLumdist} for the SM08 luminosity function, primarily the pileup feature at log(L/L$_\odot) = -4.7$, as well as the concentration at lower luminosities, present themselves in Figure \ref{sm08_lum_heatmap}. However, by showing the luminosity function with object $z$ positions, it is clear that the transition pileup at brighter luminosities does not extend as far out from the Galactic Plane as the lower luminosity object concentration; the transition pileup feature is primarily seen within 500 pc above/below the Galactic Plane and is nearly non-existent beyond 1 kpc. On the contrary, the distribution of lower luminosity objects is significantly more widespread, reaching well beyond 1 kpc above/below the Galactic Plane. In other words, \textit{the luminosity function for a subset of objects is dependent on the objects' vertical distance from the Galactic Plane.} This result has important implications for future astronomical surveys as technological advancements allow observers to probe more distant populations of ultracool dwarfs. 

\section{Discussion}\label{DISCUSSION}
In this section, we first compare our synthetic populations with one of the most complete brown dwarf sample to-date. We explore the resulting space densities and median ages of different spectral types. Then, we discuss object parameter functions from our simulation and their dependence on $\abs{z}$. Finally, we comment on the implications of our findings for future surveys.

\subsection{Comparison with Observations}
\begin{figure}
    \centering
    \includegraphics[width=\linewidth]{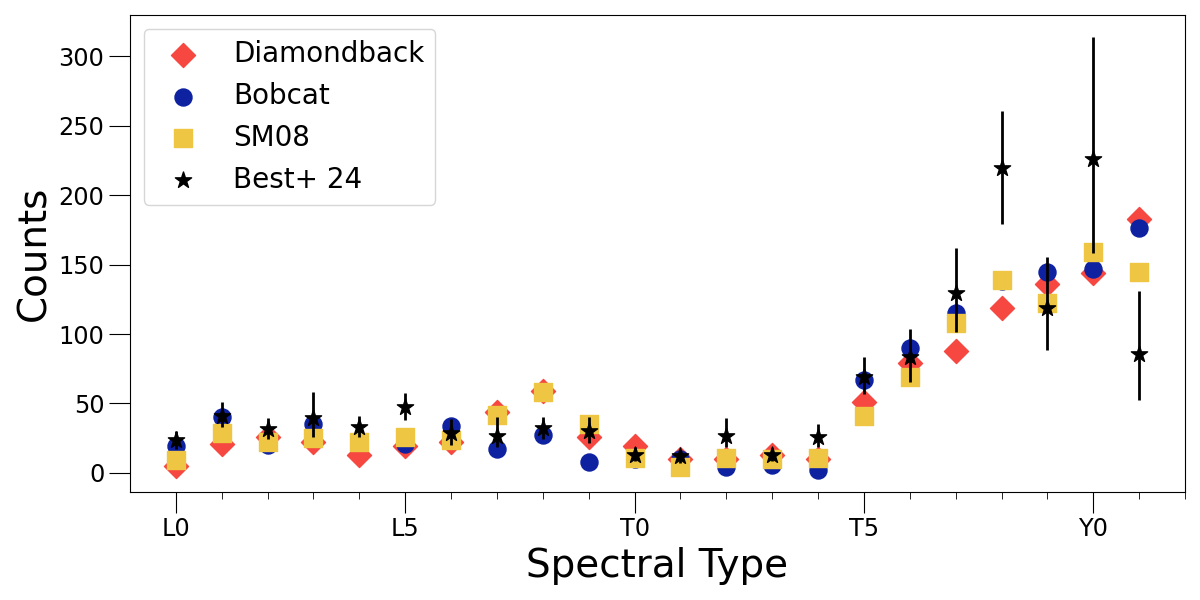}
    \caption{Simulated object counts within 25 pc by spectral type are shown in red diamonds (Sonora Diamondback), blue circles (Sonora Bobcat), and gold squares (SM08). We compare to calculated counts from \cite{Best_2024} corrected space densities with Poisson errors, denoted by black stars. All three synthetic populations are in agreement with observations for early L-type dwarfs and over-predict the late L-types before being in agreement with the observations for early T-type brown dwarfs. Conservatively, observations beyond T5 spectral types are poorly constrained and limit comparability.}
    \label{countswithin25pc}
\end{figure}

To compare our synthetic populations with local observational samples, we calculate the number of simulated objects per spectral type within 25 pc for all three synthetic populations using the effective temperature-spectral type polynomial relation from \citet{Kirkpatrick_2021_IMF}. The resulting counts are shown in Figure \ref{countswithin25pc}. For observational comparison, we use completeness-corrected space densities from \citet{Best_2024} and calculate expected counts within the 25 pc sphere. The results are shown in black with Poisson errors in Figure \ref{countswithin25pc}.  

Across all spectral types, all three model populations perform similarly, with the exception of the cloudless Sonora Bobcat population over small intervals (L6 - L8 and T6 - T7). The hybrid populations (Sonora Diamondback and SM08) are in agreement with each other for nearly all spectral types.  For early to mid L-types, the models are in agreement with observations with the exception of L5, where all three models underpredict the number of L5 dwarfs. All three models are in agreement with observations for L7 types, but struggle to match observations in later L-types. Both hybrid models overpredict L8 dwarfs while the cloudless Bobcat models underpredict L8 dwarfs, and all three models underpredict L9 types. The sudden shift from hybrid models overpredicting L8 types to underpredicting L9 types stems from the transition from cloudy to cloudless atmospheres; the SM08 models transition at 1400 K while the Sonora Diamondback models transition at 1300 K, which both occur between spectral types L7 - L9. 

Our simulations agree with observations for T dwarfs out to T7 aside from a spike in the observational counts of T2 dwarfs. All three models perform nearly identically over the T0 - T5 range; this is expected since early T dwarfs are well characterized by cloudless atmospheres. Each of the early T spectral type bins contain few ($\leq$ 15) objects. This is partly expected since the temperature-spectral type relation from \cite{Kirkpatrick_2021_IMF} is flat and only changes $\sim$100 K between L9 and T4 spectral types. For late T type dwarfs, \citet{Best_2024} urge caution as the space densities are derived from significantly incomplete samples. However, to first order, we expect the number of objects in the coldest spectral types to increase as initial mass functions favor lower mass objects and brown dwarfs continually cool, leading to more cold objects in older sub-populations. Thus, the increase in counts for synthetic populations beyond T8 dwarfs is plausible. A more complete sample of late-T and Y dwarfs is necessary to constrain the low temperature end of our population model.

\begin{figure*}
    \centering
    \includegraphics[width=\textwidth]{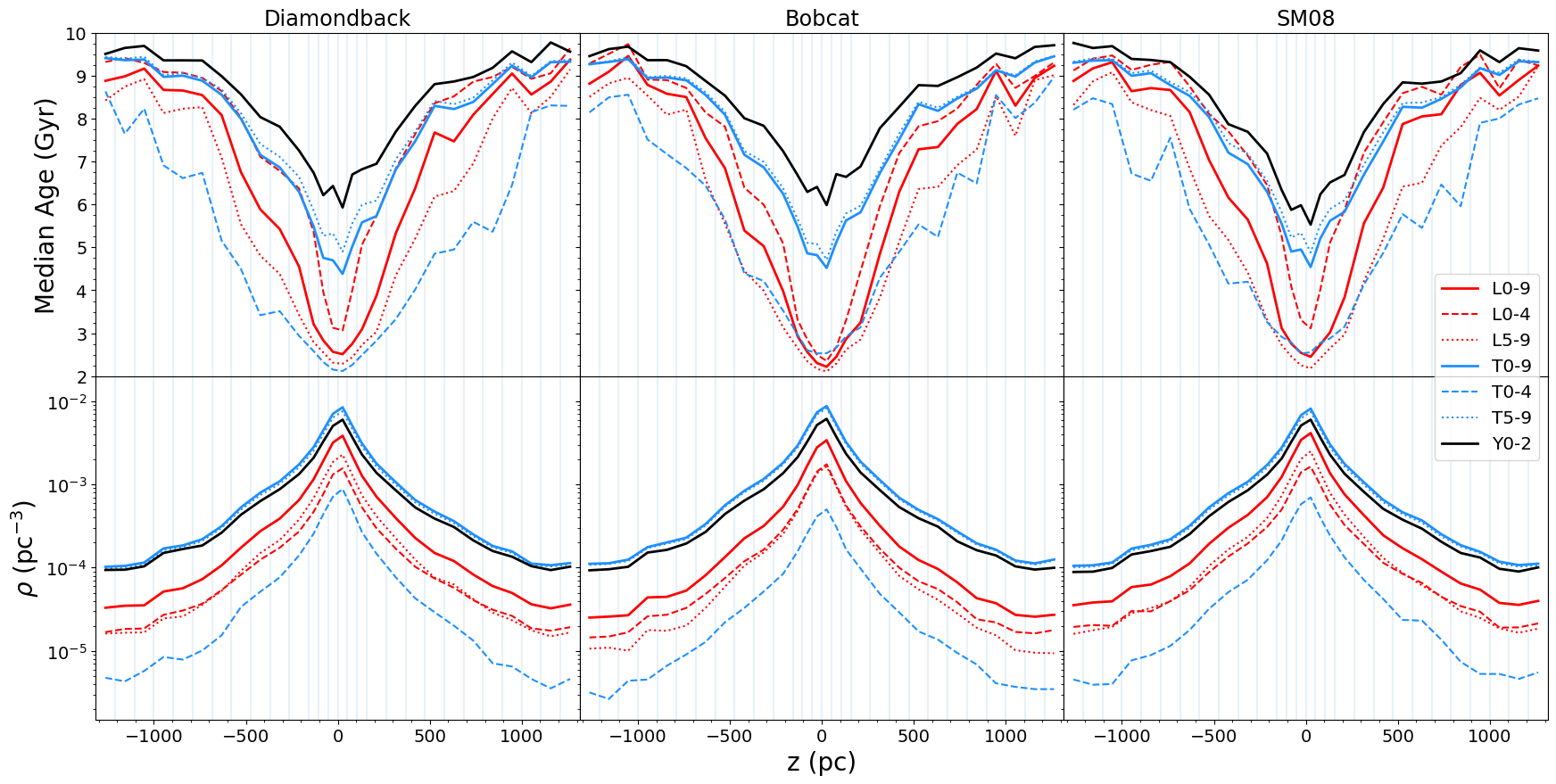}
    \caption{Median ages and space densities by spectral type by slice for all three synthetic populations. For all six panels, the vertical blue lines denote the slice edges from \citet{mazzi_SFR_2024}. Spectral types are red for L dwarfs, blue for T dwarfs, and black for Y dwarfs. Early spectral types (0 - 4) are represented by dashed lines and late spectral types (5 - 9)  are shown by dotted lines. For all models, Y dwarfs and late-T dwarfs are the oldest objects and have the highest space densities across all slices.}
    \label{medianages_spdensities}
\end{figure*}

\begin{figure*}
    \centering
    \includegraphics[width = \textwidth]{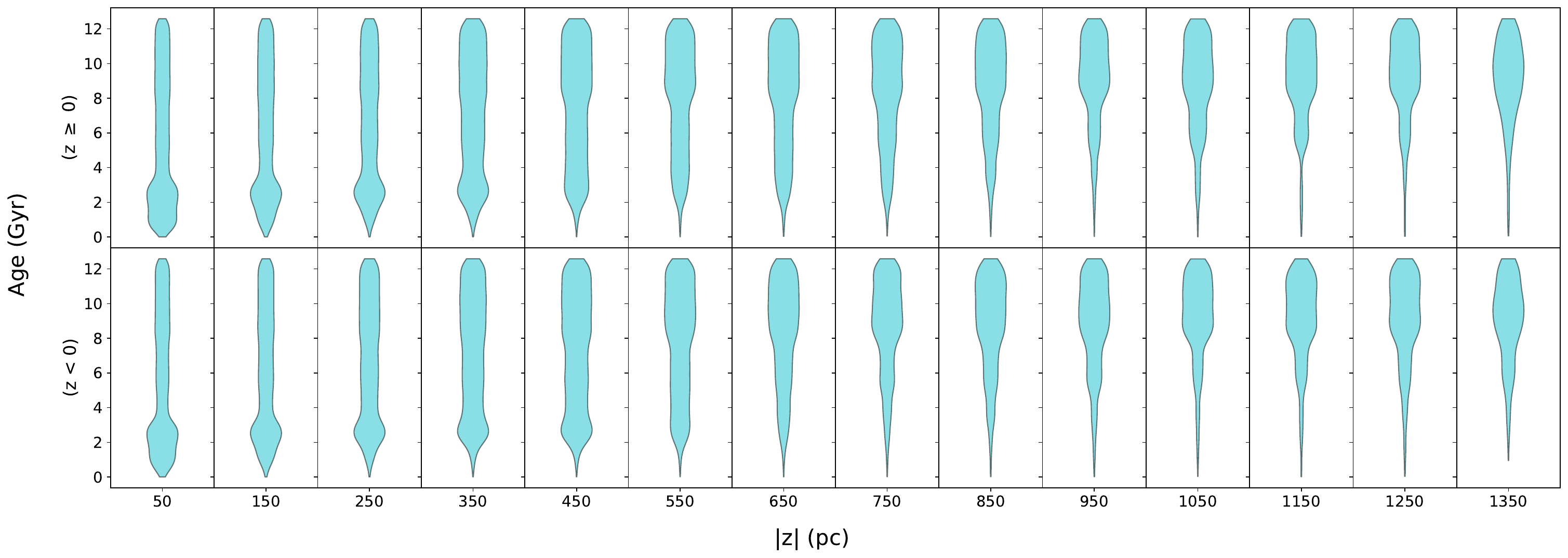}
    \caption{The underlying, SFR-based age distribution in 100 pc intervals. The upper row of panels correspond to positions above the Plane while the lower row corresponds to positions below the Plane. The age distributions of objects (and by extension the SFR history) is nearly symmetric above and below the Plane, but contains some asymmetries directly stemming from the Gaia-based SFR. Physical slices closest to the Plane are dominated by younger ages whereas slices farther from the Plane contain progressively older populations.}
    \label{violin_age}    
\end{figure*}

\subsubsection{Space Densities \& Scale Heights} \label{scaleheights}

With the calculated spectral types, we also explore the space densities and median ages by spectral type for each of the three populations. The resulting median ages are shown in the upper panels of Figure \ref{medianages_spdensities} and the space densities are shown in the lower panels. The results are also tabulated in Table \ref{spacedensities_table} and Table \ref{medianages_table} for the Diamondback population space densities and median ages, respectively\footnote{Analogous tables for the Bobcat and SM08 populations are available in the Appendix.}. We discuss median ages in depth in Section \ref{medages}.

For all three models, the space densities are highest for late-T dwarfs (T5 - T9), then Y dwarfs, L dwarfs, and finally early-T dwarfs (T0 - T4). The early and late L dwarf space densities are comparable for all three models, but early or late types do not consistently have higher or lower space densities than each other. The early-T dwarf space densities are significantly lower than other spectral types for all models across the entire volume, again stemming from the flattening of the temperature-spectral type relation.

We compute scale heights for the synthetic populations by fitting the spectral type space densities to a double exponential disk model. We only consider a thin and thick disk component, and we do not fit to a halo model since the underlying SFR history from \citet{mazzi_SFR_2024} assumes a constant halo contribution and not a flat spheroidal model like in \citet{juric08}, for example. For a single disk model, we adopt an exponential disk model as done in \citet{juric08} and \citet{Aganze_scales}:
\begin{equation}
    \rho(\textbf{r}) = \rho(R,Z) = \rho_\odot \exp(-\frac{R - R_\odot}{L})\exp(-\frac{|Z-Z_\odot|}{H}),
\end{equation}
where $\rho_\odot$ is the space density near the Sun and $R$ and $Z$ are galactocentric cylindrical coordinates such that $R^2 = X^2 + Y^2$. $L$ and $H$ are the radial and vertical scale heights, respectively, where $L=2600$ pc \citep{juric08} and we fit for $H$. However, since our simulation only extends 200 pc radially from the Sun, we approximate the radial term to be $\approx 1$. This simplification is further warranted given that we calculate space densities over each slice's volume, and therefore the space densities are at $R = R_\odot$. Thus, our simplified single disk model becomes,
\begin{equation}
    \rho(\textbf{r}) = \rho(Z) = \rho_\odot\exp(-\frac{|Z-Z_\odot|}{H}),
\end{equation}
where we only consider deviations in the vertical direction. To model both the thin and thick disk components, we use the same single component model twice, scaling the thick disk component by a fraction, $f$, and allowing each disk to have its own characteristic vertical scale height, $H_1$ and $H_2$, for the thin and thick disks, respectively:
\begin{equation}
    \rho(Z) = \rho_\odot[\exp(-\frac{|Z-Z_\odot|}{H_1}) + f\exp(-\frac{|Z-Z_\odot|}{H_2})].
\end{equation}
We fit our space densities for each synthetic population and spectral type range (L0-9, T0-9, and Y0-2) to the double disk model to obtain $H_1$, $f$, and $H_2$ values. However, for all three synthetic populations, the resulting thick disk scale heights were unreliable with extremely large uncertainties since the scale heights were on the order of the simulation vertical size. As such, we do not report $H_2$ fit values and instead assume a fixed thick disk scale height of $900$ pc \citep{juric08}.  We present fit values for $H_1$ and $f$ for all three evolutionary models and L, T, and Y dwarfs separately in Table \ref{scale heights}. 

In all three cases, our fit thin disk scale height is significantly lower than published values of $\sim$ 300 pc \citep{juric08}. As discussed in \citet{mazzi_SFR_2024}, the Galactic disk scale heights are a function of age; younger populations of objects will yield smaller vertical scale heights while older populations that have had time to dynamically scatter yield larger vertical scale heights. Given the young age distributions of our objects, particularly compared to older stellar populations, shorter vertical scale heights are not unexpected.

\begin{deluxetable}{c c c c}
    \tablecolumns{4}
    \label{scale heights}
    \tablecaption{Scale height parameter fits by spectral type from double exponential disk model.}
    \tablehead{\colhead{Model} & \colhead{Sp. Type} & \colhead{$H_1$}  & \colhead{$f$}\\
    \colhead{}  & \colhead{} & \colhead{(pc)} & \colhead{($\times10^{-2}$)}}
    \startdata
    Diamondback & L & $115\pm10$ & $4.0\pm2.5$ \\
    {} & T & $122\pm11$ & $5.3\pm2.4$ \\
    {} & Y & $125\pm11$ & $6.2\pm2.5$ \\
    \hline
    Bobcat & L & $113\pm10$ & $3.6\pm2.4$ \\
    {} & T & $123\pm11$ & $5.4\pm2.5$ \\
    {} & Y & $124\pm11$ & $6.2\pm2.5$ \\
    \hline
    SM08 & L & $114\pm10$ & $4.1\pm2.5$ \\
    {} & T & $123\pm10$ & $5.4\pm2.4$ \\
    {} & Y & $124\pm11$ & $6.0\pm2.5$ \\
    \enddata
    
\end{deluxetable}

\subsubsection{Median Ages}\label{medages}

The Sonora Diamondback, Sonora Bobcat, and SM08 models all independently calculate evolutionary tracks (with the exception of Diamondback, which reverts to the Bobcat evolutionary tracks below 900 K, equivalent to T6 spectral types). As such, while the median ages of late T dwarfs and Y dwarfs appears consistent across models for all slices, the median ages of L dwarfs and early-T dwarfs is distinctly different for each model. In all three populations, early-L dwarfs (L0 - L4) are older than late-L dwarfs (L5 - L9). The earlier L dwarfs have an older median age than late-L dwarfs since, in addition to young brown dwarfs, this spectral range includes older, more massive, stellar populations that have cooled sufficiently to reach the L-dwarf temperature range. On the other hand, the L5 - L9 spectral type range is exclusively comprised of brown dwarfs and does not include an older ultracool dwarf component. The primary difference between the three models is the median age for L dwarfs, particularly in comparison to early-T dwarfs near the Plane, where objects are more numerous and younger overall. The differences across the three models can be attributed to the different cooling mechanisms and timescales associated with each model. The Bobcat model, which does not include an L-T transition, yields nearly identical L dwarf and early-T dwarf median ages in the solar slice with late-L dwarfs actually being older than early-T dwarfs. SM08 predicts the same age comparison while Diamondback predicts late-L dwarfs to be younger than early-T dwarfs. These differences underscore the different cooling timescales for the gravity-dependent transition in Diamondback, linear transition in SM08, and non-transition in Bobcat.

\begin{deluxetable*}{c | c c | c c | c | c }
    \tablecolumns{7}
    \label{agecomp}
    \tablecaption{Comparison of median ages with literature.}
    \tablehead{\colhead{Source} & \colhead{L0-4} & \colhead{L5-9}  & \colhead{T0-4} & \colhead{T5-9} & \colhead{L0-9} & \colhead{T0-9}\\
    \colhead{}  & \colhead{(Gyr)} & \colhead{(Gyr)} & \colhead{(Gyr)} & \colhead{(Gyr)} & \colhead{(Gyr)} & \colhead{(Gyr)} }
    \startdata
    Diamondback\tablenotemark{a} & 5.58 & 2.92 & 2.60 & 6.41 & 3.63 & 6.01  \\
    Bobcat\tablenotemark{a} & 3.89 & 2.72 & 3.03 & 6.25 & 3.05 & 6.07\\
    SM08\tablenotemark{a} & 5.76 & 2.85 & 3.06 & 6.41 & 3.61 & 6.12\\
    \hline
    Diamondback\tablenotemark{b} & 3.06 & 2.29 & 2.12 & 4.90 & 2.51 & 4.38 \\
    Bobcat\tablenotemark{b} & 2.35 & 2.11 & 2.53 & 4.71 & 2.22 & 4.52\\
    SM08\tablenotemark{b} & 3.11 & 2.18 & 2.56 & 4.88 & 2.45 & 4.54 \\
    \hline
    \citet{Aganze_scales}  & $2.0^{+0.6}_{-0.4}$ & $2.4^{+2.9}_{-0.8}$ & $2.5^{+2.8}_{-0.9}$ & $2.6^{+3.9}_{-1.0}$ & $2.1^{+0.9}_{-0.5}$ & $2.4^{+2.4}_{-0.8}$\\
    \citet{burgasser15}            & $6.5\pm0.4$ & -- & -- & -- & -- & -- \\
    \citet{hsu21}                 & -- & -- & -- & -- & $4.2\pm0.3$ & $3.5\pm0.3$ \\
    %\citet{Faherty_BDKP1} & -- & -- & -- & -- & $2.6^{+0.9}_{-0.7}$ & $2.2^{+0.8}_{-0.7}$\\
    \citet{dupuy17} & -- & -- & -- & --  &1.3\tablenotemark{c} & --\\
    \enddata
    \tablecomments{\tablenotetext{a}{Full simulated volume}\tablenotetext{b}{Solar slice only ($0.00 \leq z < 52.63$ pc)} \tablenotetext{c}{The \citet{dupuy17} sample contains M7 - T5 spectral types, not only L0-9}}
\end{deluxetable*}

For the solar slice, highlighted in Tables \ref{spacedensities_table} \& \ref{medianages_table}, we compare our median ages to local samples. For early-L dwarfs, \citet{Aganze_scales} found a median age of $2.0^{+0.6}_{-0.4}$ Gyr and \citet{burgasser15} found $6.5\pm0.4$ Gyr.  For late-L dwarfs, \citet{Aganze_scales} found a median age of $2.4^{+2.9}_{-0.8}$ Gyr. \citet{hsu21} found $4.2\pm0.3$ Gyr for L dwarfs overall, and \citet{Aganze_scales} found $2.1^{+0.9}_{-0.5}$. For T dwarfs, \citet{Aganze_scales} found a median age of $2.4^{+2.4}_{-0.8}$ Gyr and \citet{hsu21} reports $3.5\pm0.3$ Gyr. Within the solar slice, our simulated median ages for all three synthetic populations are in agreement with values and uncertainties reported in \citet{Aganze_scales}, with the exception of the Diamondback and SM08 L0-4 ages. Our L0-4 ages are older due to a stellar component of objects in our simulation arising from the age and mass ranges used as inputs.

To compare our median age to the 1.3 Gyr median age reported in \citet{dupuy17}, we perform $10^6$ Monte Carlo trials. Their sample contained 10 binary systems spanning M7 - T5 spectral types with distances between 10 and 40 pc. For each trial, we draw 10 objects from our synthetic population following the spectral type distribution in \citet{dupuy17}\footnote{\citet{dupuy17} reported the breakdown of spectral types in their age-dated sample was 28\% L6 or earlier, 33\% L6.5 - L8.5, and 39\% L9 - T5. Subsequently, in our Monte Carlo trials, we select 3 objects from L0 - L6, 3 objects from L7 - L8, and 4 objects from L9 - T5.} and with distances between 10 and 40 pc.

We found the median age for the Diamondback population to be $2.7$ Gyr, with 90\% of the distribution contained between $1.3 - 4.7$ Gyr; for the Bobcat population, the median age was $2.9$ Gyr and 90\% of the distribution was contained between $1.5 - 5.4$ Gyr; finally, for the SM08 population, the median age was $2.7$ Gyr and 90\% of the distribution was contained between $1.3 - 5.2$ Gyr. In all three populations, less than 5\% of trials yield median ages less than or equal to 1.3 Gyr. All three populations yield older median ages than the \citet{dupuy17} binary sample, but are roughly in agreement with the binary sample's mean age of $2.3$ Gyr. Kinematically, the median and dispersion of the tangential velocity are higher in the volume-complete \citet{kirkpatrick24} sample than in the \citet{dupuy17} binary sample, suggesting a possible bias towards younger ages \citep{Wielen74, Wielen77, Binney09}.

\startlongtable
\begin{deluxetable*}{ c c|c c c c|c c c|c}
\tablecolumns{10}
\label{spacedensities_table}
\tablecaption{Space densities, $\rho$, by spectral type within \textit{z} slices for the Sonora Diamondback synthetic population. All space densities are reported in ($\times 10^{-3}$ pc$^{-3}$).}
\tablehead{\colhead{z$_{\text{min}}$} & \colhead{z$_{\text{max}}$} & \colhead{L0-4} & \colhead{L5-9} & \colhead{T0-4} & \colhead{T5-9} & \colhead{L0-9} & \colhead{T0-9} & \colhead{Y0-2} & \colhead{Total}\\
\colhead{(pc)} & \colhead{(pc)} & \colhead{} & \colhead{} & \colhead{} & \colhead{} & \colhead{} & \colhead{} & \colhead{} & \colhead{} }
\startdata 
$-1315.78$ & $-1210.52$ & 0.017 & 0.016 & 0.005 & 0.098 & 0.033 & 0.103 & 0.094 & 0.230 \\
$-1210.52$ & $-1105.26$ & 0.018 & 0.017 & 0.004 & 0.101 & 0.035 & 0.105 & 0.095 & 0.235 \\
$-1105.26 $& $-1000.00$ & 0.019 & 0.017 & 0.006 & 0.109 & 0.035 & 0.115 & 0.104 & 0.254 \\
$-1000.00$ & $-894.74$ & 0.027 & 0.025 & 0.008 & 0.162 & 0.052 & 0.170 & 0.150 & 0.372 \\
$-894.74$ & $-789.47$ & 0.031 & 0.026 & 0.008 & 0.177 & 0.057 & 0.185 & 0.168 & 0.410 \\
$-789.47$ & $-684.21$ & 0.037 & 0.036 & 0.010 & 0.210 & 0.073 & 0.220 & 0.185 & 0.477 \\
$-684.21$ & $-578.95$ & 0.054 & 0.053 & 0.015 & 0.300 & 0.107 & 0.315 & 0.263 & 0.685 \\
$-578.95$ & $-473.68$ & 0.083 & 0.092 & 0.034 & 0.497 & 0.174 & 0.531 & 0.433 & 1.138 \\
$-473.68$ & $-368.42$ & 0.125 & 0.150 & 0.052 & 0.746 & 0.275 & 0.798 & 0.631 & 1.704 \\
$-368.42$ & $-263.16$ & 0.174 & 0.215 & 0.076 & 1.014 & 0.388 & 1.091 & 0.877 & 2.356 \\
$-263.16$ & $-157.89$ & 0.272 & 0.378 & 0.140 & 1.603 & 0.651 & 1.743 & 1.332 & 3.725 \\
$-157.89$ & $-105.26$ & 0.468 & 0.683 & 0.256 & 2.546 & 1.151 & 2.802 & 2.089 & 6.042 \\
$-105.26$ & $-52.63$ & 0.768 & 1.149 & 0.440 & 4.023 & 1.916 & 4.463 & 3.281 & 9.660 \\
$-52.63$ & 0.00 & 1.319 & 1.862 & 0.713 & 6.355 & 3.181 & 7.069 & 5.052 & 15.302 \\
\hline
0.00 & 52.63 & 1.563 & 2.285 & 0.882 & 7.569 & 3.848 & 8.451 & 6.001 & 18.300 \\
\hline
52.63 & 105.26 & 0.904 & 1.295 & 0.499 & 4.579 & 2.199 & 5.079 & 3.689 & 10.967 \\
105.26 & 157.89 & 0.530 & 0.739 & 0.267 & 2.845 & 1.269 & 3.112 & 2.288 & 6.668 \\
157.89 & 263.16 & 0.298 & 0.415 & 0.147 & 1.635 & 0.712 & 1.783 & 1.379 & 3.873 \\
263.16 & 368.42 & 0.169 & 0.225 & 0.076 & 0.985 & 0.394 & 1.061 & 0.849 & 2.304 \\
368.42 & 473.68 & 0.103 & 0.125 & 0.043 & 0.608 & 0.228 & 0.650 & 0.529 & 1.407 \\
473.68 & 578.95 & 0.074 & 0.076 & 0.029 & 0.443 & 0.150 & 0.472 & 0.389 & 1.011 \\
578.95 & 684.21 & 0.057 & 0.063 & 0.020 & 0.342 & 0.120 & 0.362 & 0.308 & 0.791 \\
684.21 & 789.47 & 0.041 & 0.042 & 0.013 & 0.238 & 0.083 & 0.251 & 0.213 & 0.547 \\
789.47 & 894.74 & 0.031 & 0.029 & 0.007 & 0.175 & 0.060 & 0.183 & 0.159 & 0.401 \\
894.74 & 1000.00 & 0.026 & 0.023 & 0.007 & 0.151 & 0.050 & 0.157 & 0.135 & 0.342 \\
1000.00 & 1105.26 & 0.019 & 0.018 & 0.005 & 0.107 & 0.036 & 0.112 & 0.104 & 0.253 \\
1105.26 & 1210.52 & 0.018 & 0.015 & 0.004 & 0.104 & 0.033 & 0.107 & 0.094 & 0.233 \\
1210.52 & 1315.78 & 0.019 & 0.017 & 0.005 & 0.109 & 0.036 & 0.114 & 0.103 & 0.253 \\
\enddata
\end{deluxetable*}

\startlongtable
\begin{deluxetable*}{ c c | c c c c | c c c }
\tablecolumns{9}
\label{medianages_table}
\tablecaption{Median ages by spectral type within \textit{z} slices for the Sonora Diamondback population.}
\tablehead{\colhead{z$_{\text{min}}$} & \colhead{z$_{\text{max}}$} & \colhead{L0-4} & \colhead{L5-9} & \colhead{T0-4} & \colhead{T5-9} & \colhead{L0-9} & \colhead{T0-9} & \colhead{Y0-2} \\
\colhead{(pc)} & \colhead{(pc)} & \colhead{(Gyr)} & \colhead{(Gyr)} & \colhead{(Gyr)} & \colhead{(Gyr)} & \colhead{(Gyr)} & \colhead{(Gyr)} & \colhead{(Gyr)} }
\startdata
-1315.78 & -1210.52 & 9.32 & 8.42 & 8.63 & 9.44 & 8.88 & 9.41 & 9.51 \\
-1210.52 & -1105.26 & 9.41 & 8.74 & 7.65 & 9.40 & 8.98 & 9.36 & 9.65 \\
-1105.26 & -1000.00 & 9.30 & 8.92 & 8.23 & 9.44 & 9.17 & 9.38 & 9.69 \\
-1000.00 & -894.74 & 9.09 & 8.13 & 6.91 & 9.02 & 8.67 & 8.97 & 9.36 \\
-894.74 & -789.47 & 9.07 & 8.23 & 6.61 & 9.06 & 8.65 & 9.00 & 9.36 \\
-789.47 & -684.21 & 8.95 & 8.27 & 6.74 & 8.95 & 8.55 & 8.88 & 9.35 \\
-684.21 & -578.95 & 8.64 & 7.09 & 5.16 & 8.68 & 8.08 & 8.55 & 8.99 \\
-578.95 & -473.68 & 8.02 & 5.54 & 4.50 & 8.12 & 6.75 & 7.99 & 8.57 \\
-473.68 & -368.42 & 7.11 & 4.81 & 3.42 & 7.41 & 5.88 & 7.16 & 8.03 \\
-368.42 & -263.16 & 6.78 & 4.38 & 3.51 & 7.11 & 5.43 & 6.86 & 7.81 \\
-263.16 & -157.89 & 6.37 & 3.43 & 2.94 & 6.64 & 4.55 & 6.29 & 7.25 \\
-157.89 & -105.26 & 5.37 & 2.80 & 2.58 & 5.86 & 3.21 & 5.49 & 6.74 \\
-105.26 & -52.63 & 3.96 & 2.55 & 2.33 & 5.27 & 2.83 & 4.76 & 6.22 \\
-52.63 & 0.00 & 3.13 & 2.31 & 2.15 & 5.33 & 2.57 & 4.70 & 6.43 \\
\hline
0.00 & 52.63 & 3.06 & 2.29 & 2.12 & 4.90 & 2.51 & 4.38 & 5.93 \\
\hline
52.63 & 105.26 & 4.00 & 2.44 & 2.26 & 5.56 & 2.76 & 5.02 & 6.70 \\
105.26 & 157.89 & 5.06 & 2.73 & 2.49 & 5.99 & 3.09 & 5.58 & 6.82 \\
157.89 & 263.16 & 5.74 & 3.03 & 2.83 & 6.09 & 3.87 & 5.73 & 6.95 \\
263.16 & 368.42 & 6.84 & 4.35 & 3.33 & 7.05 & 5.33 & 6.82 & 7.70 \\
368.42 & 473.68 & 7.64 & 5.21 & 4.02 & 7.70 & 6.38 & 7.47 & 8.30 \\
473.68 & 578.95 & 8.36 & 6.18 & 4.85 & 8.41 & 7.68 & 8.30 & 8.80 \\
578.95 & 684.21 & 8.52 & 6.31 & 4.95 & 8.33 & 7.47 & 8.22 & 8.86 \\
684.21 & 789.47 & 8.86 & 6.96 & 5.59 & 8.49 & 8.09 & 8.39 & 8.97 \\
789.47 & 894.74 & 8.97 & 8.01 & 5.36 & 8.89 & 8.57 & 8.80 & 9.18 \\
894.74 & 1000.00 & 9.20 & 8.72 & 6.45 & 9.31 & 9.05 & 9.23 & 9.57 \\
1000.00 & 1105.26 & 8.93 & 8.13 & 8.15 & 8.99 & 8.56 & 8.96 & 9.32 \\
1105.26 & 1210.52 & 9.06 & 8.51 & 8.31 & 9.34 & 8.86 & 9.31 & 9.77 \\
1210.52 & 1315.78 & 9.64 & 9.15 & 8.30 & 9.37 & 9.37 & 9.33 & 9.56 \\
\enddata
\end{deluxetable*}

\subsection{Fundamental Parameter Functions}
As future observatories, such as JWST, Euclid, Rubin, and Roman, commence operations and detect brown dwarfs all the way into the Galactic Halo, the effects of Galactic dynamics can no longer be ignored.  For small, Solar Neighborhood samples, the overall kinematics and observational effects from the Galaxy as a whole could be minimized, but such an approach will no longer be possible for larger, more distant samples. 

Our simulation presents the current state of brown dwarfs within 200 pc and 1.3 kpc above/below the Galactic Plane. Whereas local samples are inherently biased towards young objects, since older objects have higher velocities from dynamical heating \citep{Kirkpatrick_2021_IMF}, our simulation inherently accounts for these kinematics by using the Gaia-derived SFR history. As such, our simulation yields the number of objects and their ages as one would find them today, accounting for Galactic effects. We capture the older populations at higher $\abs{z}$ values as well as the young populations close to the Galactic Plane. Figure \ref{violin_age} shows how the age distribution of objects changes by 100 pc slices above and below the Plane. Slices near the Galactic Plane favor younger objects, and slices farther from the Plane progressively favor older objects; the most distant slices are almost entirely comprised of old ($\geq 8$ Gyr) objects. This trend appears both above and below the Galactic Plane with only slight differences in slices at comparable $\abs{z}$ heights. Note that the final slice, $\abs{z} \geq 1.3$ kpc shows a slightly younger distribution than the previous slice, but this a numerical effect from small number statistics as the simulation does not encompass the entire 1.3 - 1.4 kpc slice but rather only reaches $\abs{z} \approx 1.32$ kpc. Only $\sim 500$ objects fall into each of the most distant, incomplete bins above/below the Plane.

\figsetstart
\figsetnum{11}
\figsettitle{Sonora Luminosity Functions by $\abs{z}$ Slices}
\figsetgrpstart
\figsetgrpnum{11.1}
\figsetgrptitle{All slices with selected pop-outs}
\figsetplot{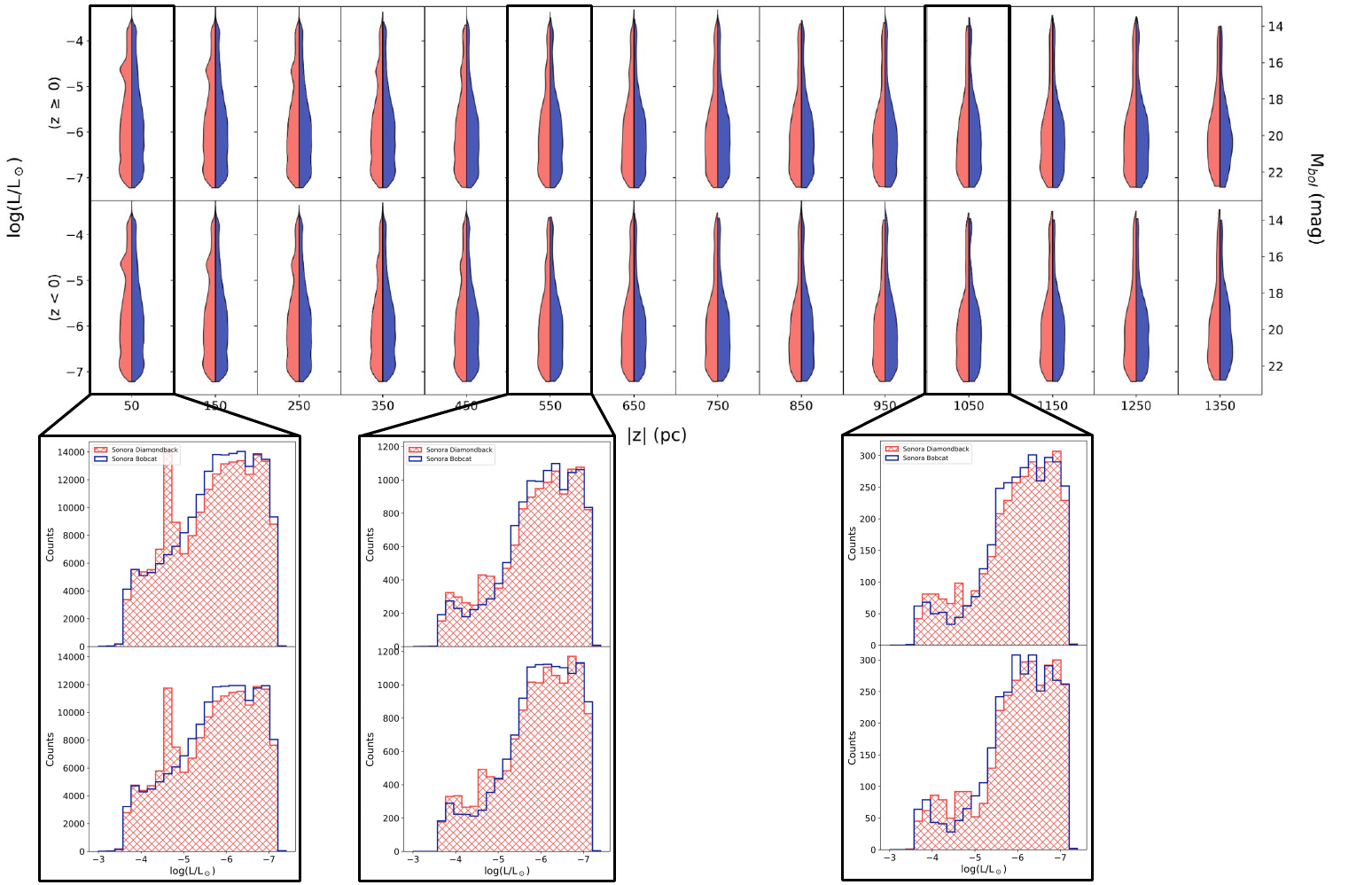}
\figsetgrpnote{The luminosity function for the Sonora Diamondback (red) and Bobcat (blue) synthetic populations is shown in 100 pc bins. As distance from the Plane increases, the cloudy-cloudless transition feature in the Diamondback model dissipates and the population shifts to the lowest luminosities. In the slices closest to the Galactic Plane, the T-Y transition pileup is visible but is indistinguishable at larger distances. The Bobcat luminosity function is smooth but extends to brighter luminosities at lower $|z|$ slices. Three slices are shown in greater detail in pop-out histograms. The histograms closest to the Galactic Plane (far left) contain more objects and show the most prominent pileup features. The middle pop-out contains fewer objects and shows the decrease in pileup feature strength. The farthest pop-out from the Plane (far right) contains the least number of objects and the weakest pileup feature strength.}
\figsetgrpend

\figsetgrpstart
\figsetgrpnum{11.2}
\figsetgrptitle{$\abs{z}$ 0-100 pc}
\figsetplot{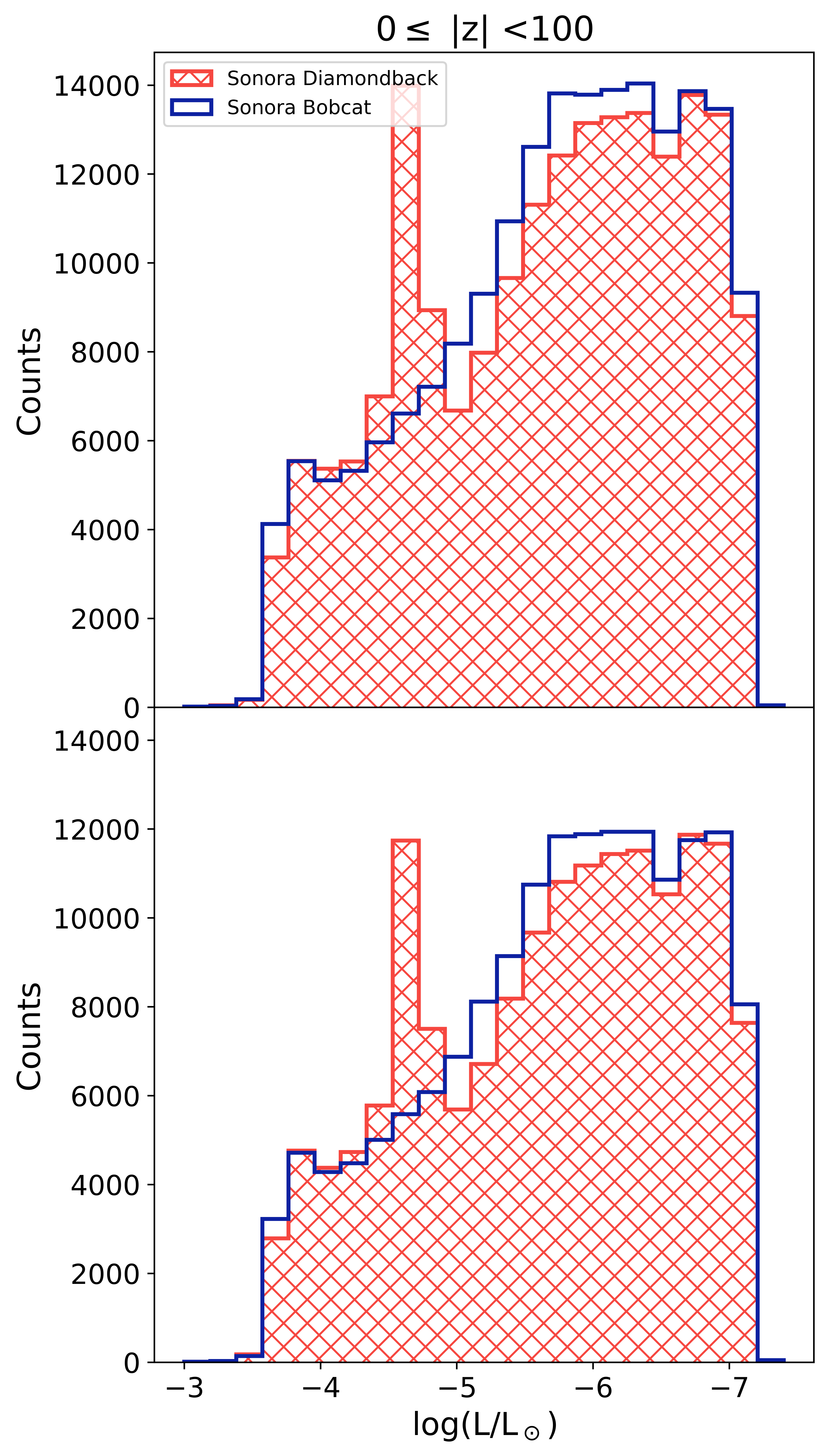}
\figsetgrpnote{Sonora luminosity functions for |z|=0-100 slice.}
\figsetgrpend

\figsetgrpstart
\figsetgrpnum{11.3}
\figsetgrptitle{$\abs{z}$ 100-200 pc}
\figsetplot{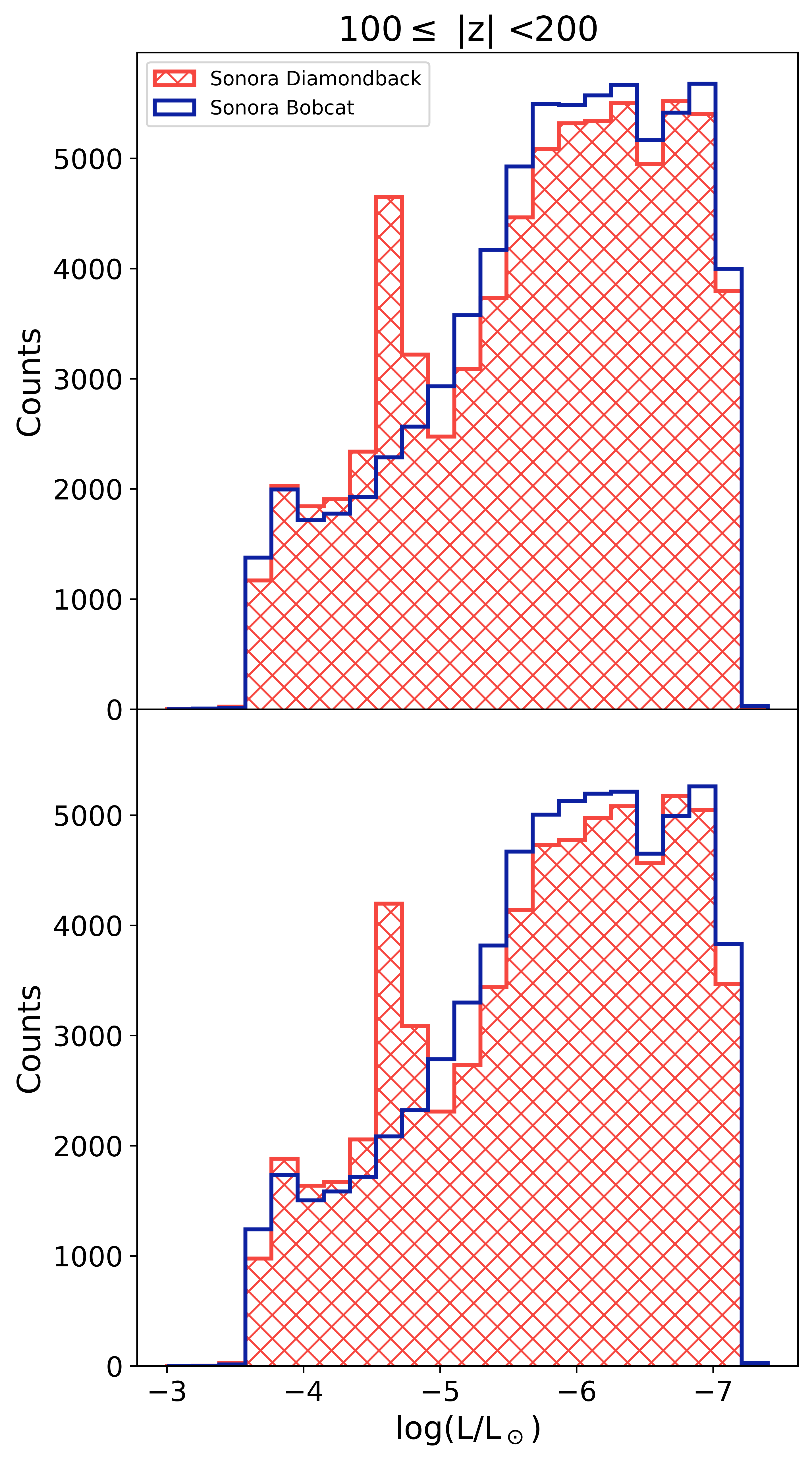}
\figsetgrpnote{Sonora luminosity functions for |z|=100-200 slice.}
\figsetgrpend

\figsetgrpstart
\figsetgrpnum{11.4}
\figsetgrptitle{$\abs{z}$ 200-300 pc}
\figsetplot{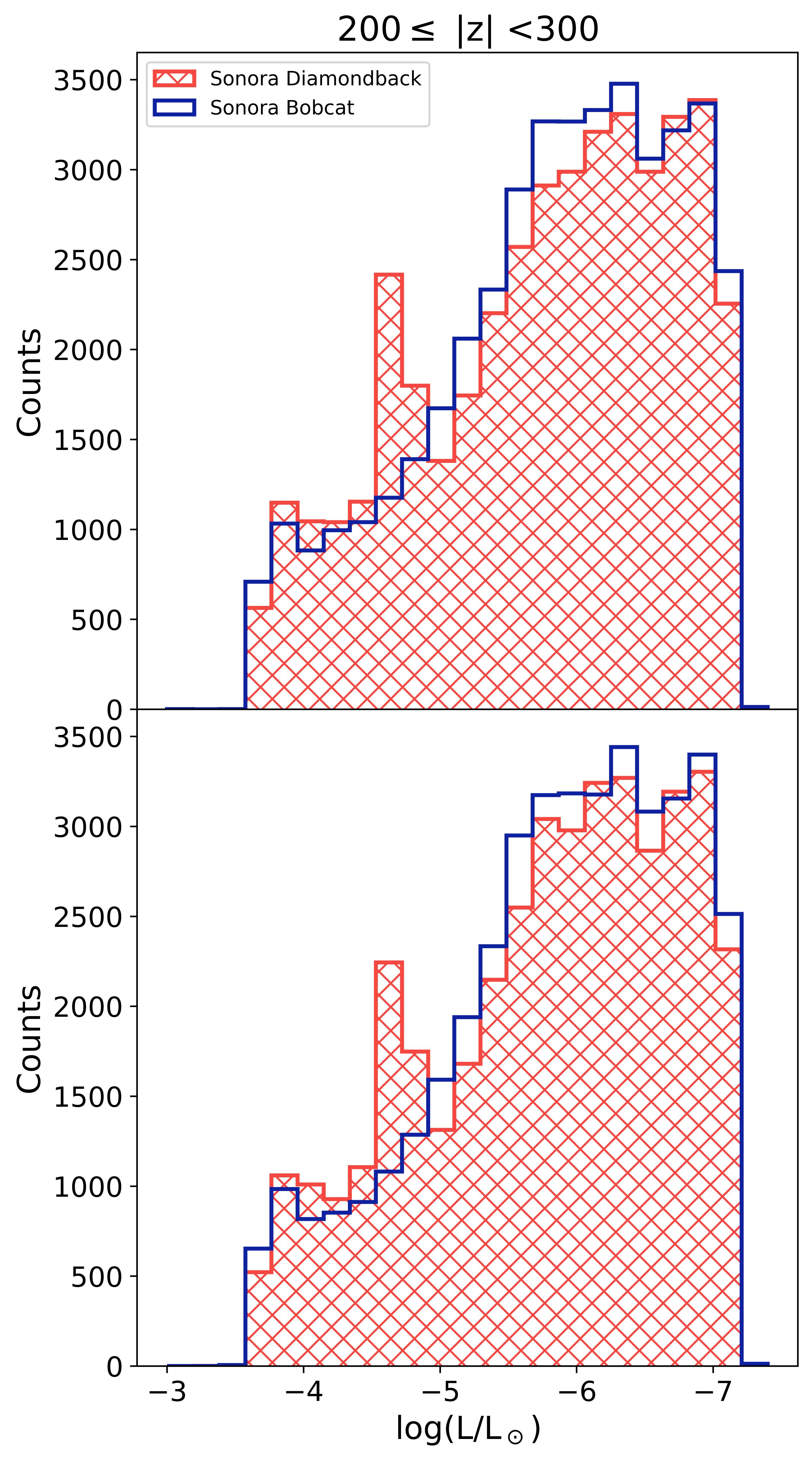}
\figsetgrpnote{Sonora luminosity functions for |z|=200-300 slice.}
\figsetgrpend

\figsetgrpstart
\figsetgrpnum{11.5}
\figsetgrptitle{$\abs{z}$ 300-400 pc}
\figsetplot{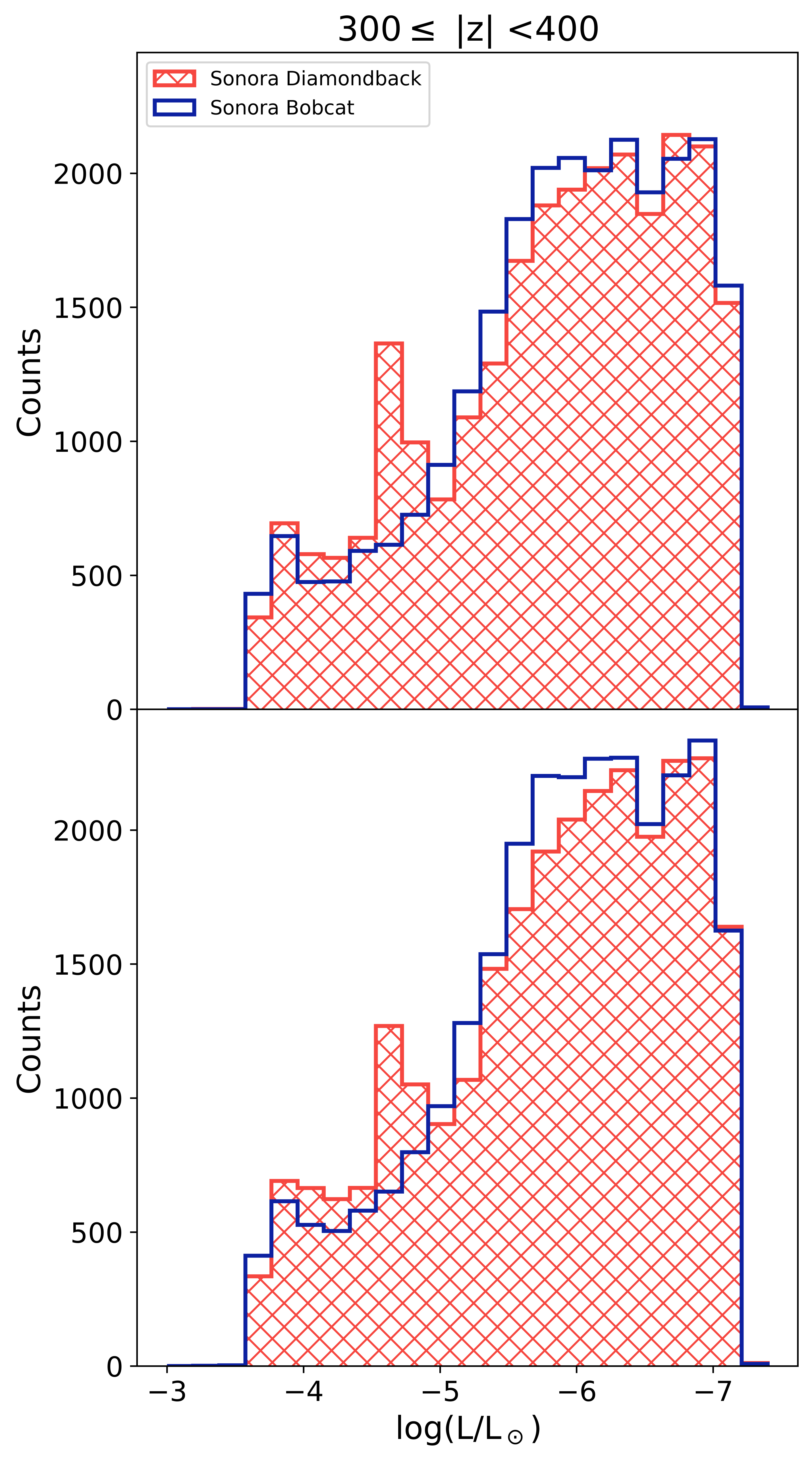}
\figsetgrpnote{Sonora luminosity functions for |z|=300-400 slice.}
\figsetgrpend

\figsetgrpstart
\figsetgrpnum{11.6}
\figsetgrptitle{$\abs{z}$ 400-500 pc}
\figsetplot{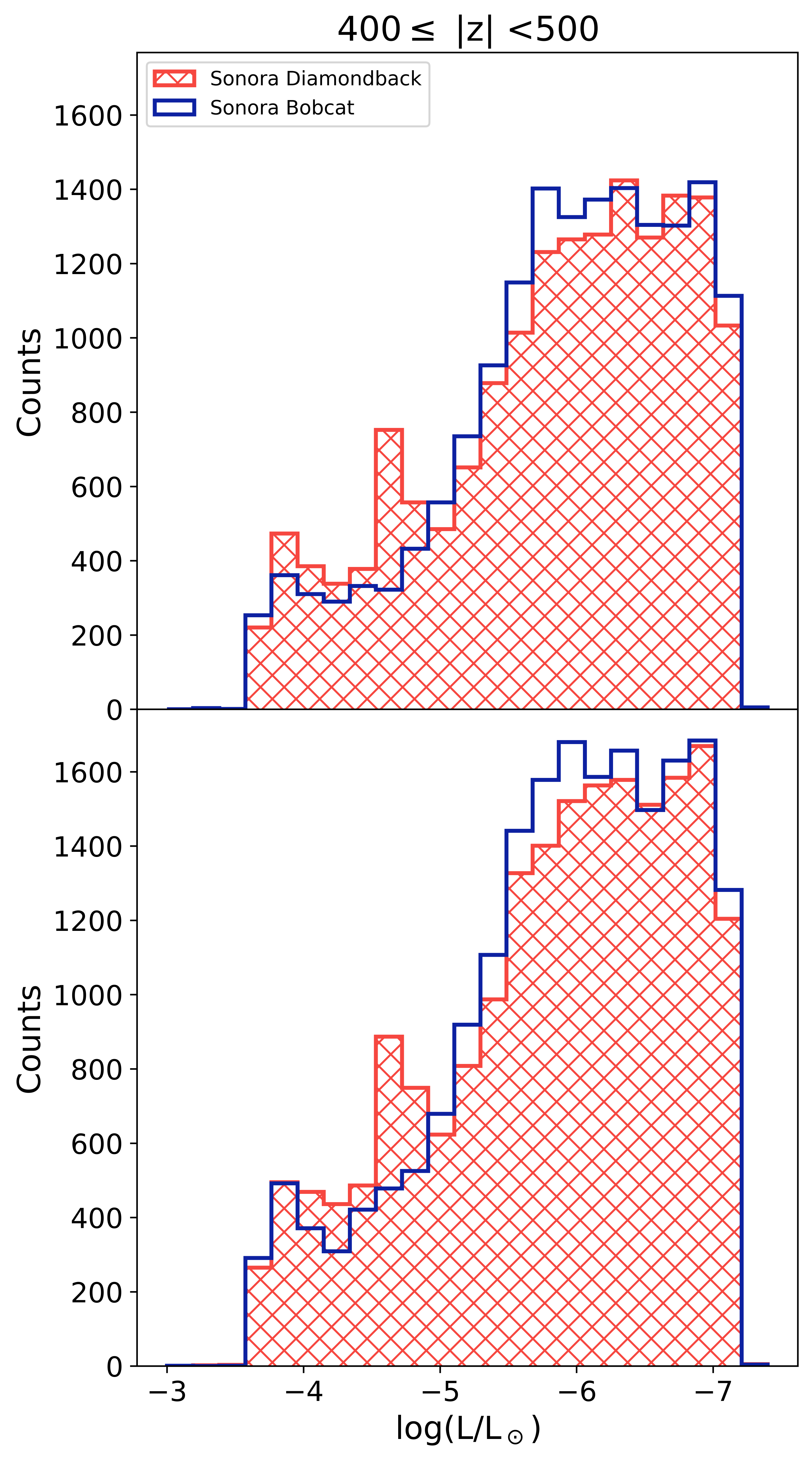}
\figsetgrpnote{Sonora luminosity functions for |z|=400-500 slice.}
\figsetgrpend

\figsetgrpstart
\figsetgrpnum{11.7}
\figsetgrptitle{$\abs{z}$ 500-600 pc}
\figsetplot{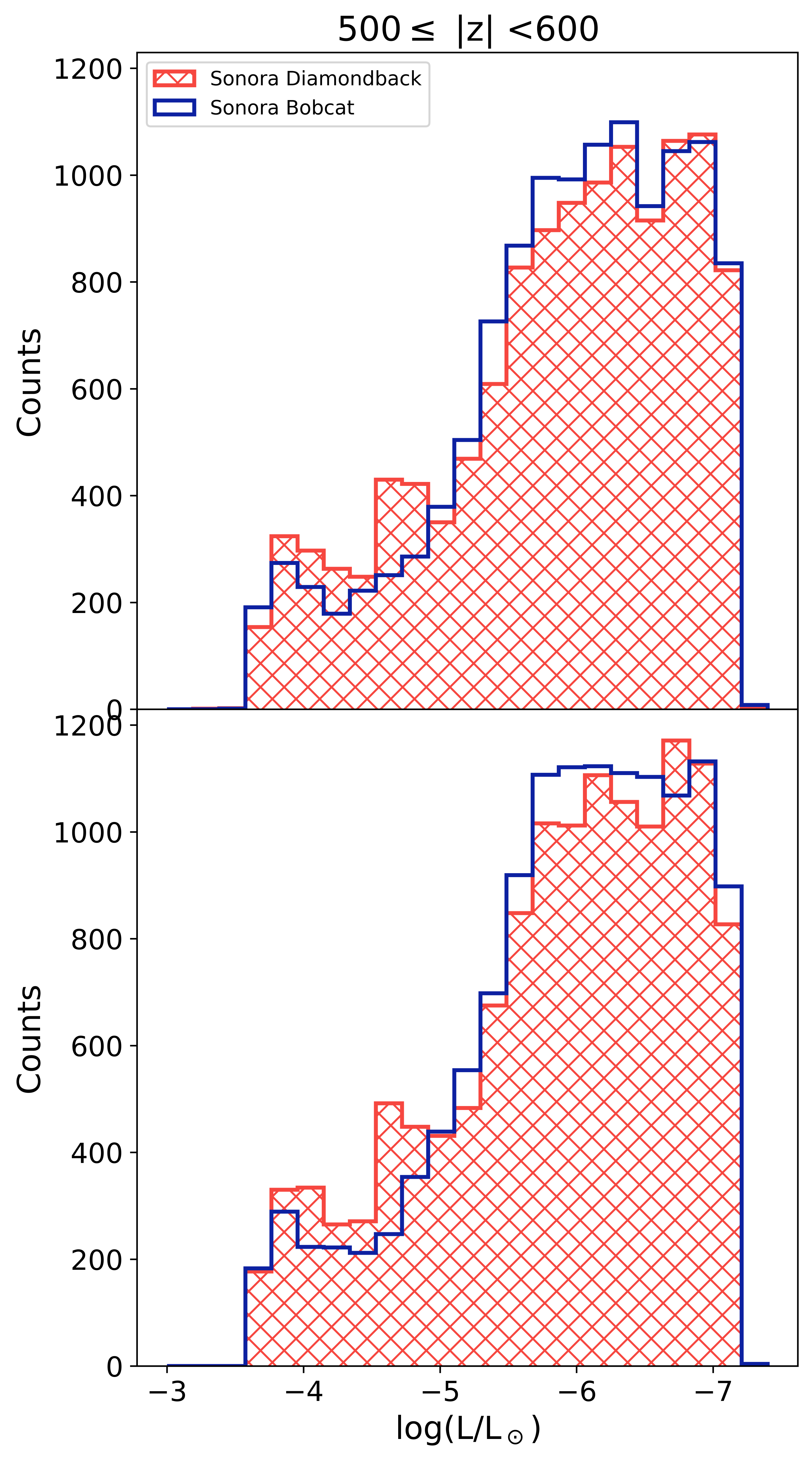}
\figsetgrpnote{Sonora luminosity functions for |z|=500-600 slice.}
\figsetgrpend

\figsetgrpstart
\figsetgrpnum{11.8}
\figsetgrptitle{$\abs{z}$ 600-700 pc}
\figsetplot{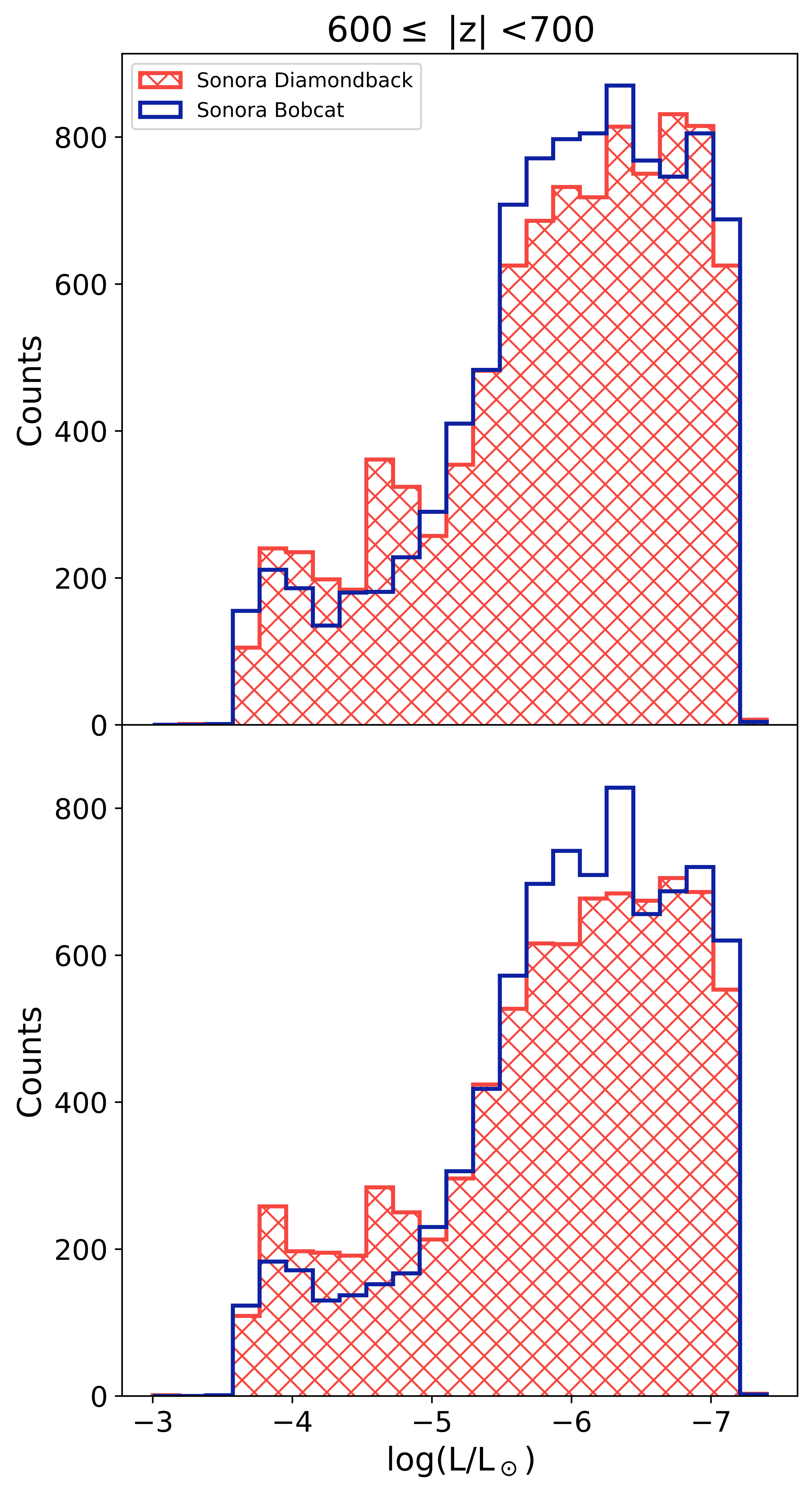}
\figsetgrpnote{Sonora luminosity functions for |z|=600-700 slice.}
\figsetgrpend

\figsetgrpstart
\figsetgrpnum{11.9}
\figsetgrptitle{$\abs{z}$ 700-800 pc}
\figsetplot{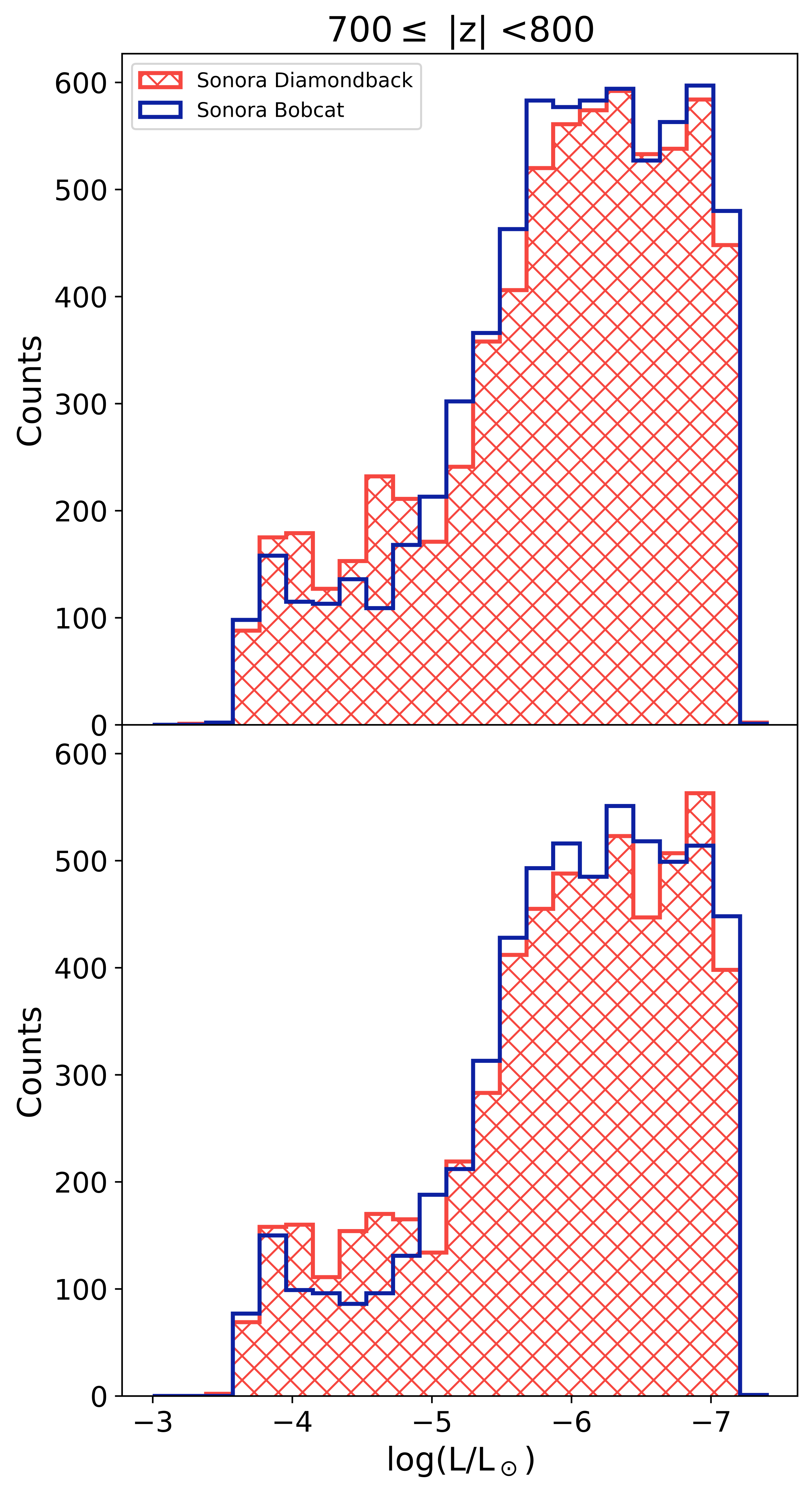}
\figsetgrpnote{Sonora luminosity functions for |z|=700-800 slice.}
\figsetgrpend

\figsetgrpstart
\figsetgrpnum{11.10}
\figsetgrptitle{$\abs{z}$ 800-900 pc}
\figsetplot{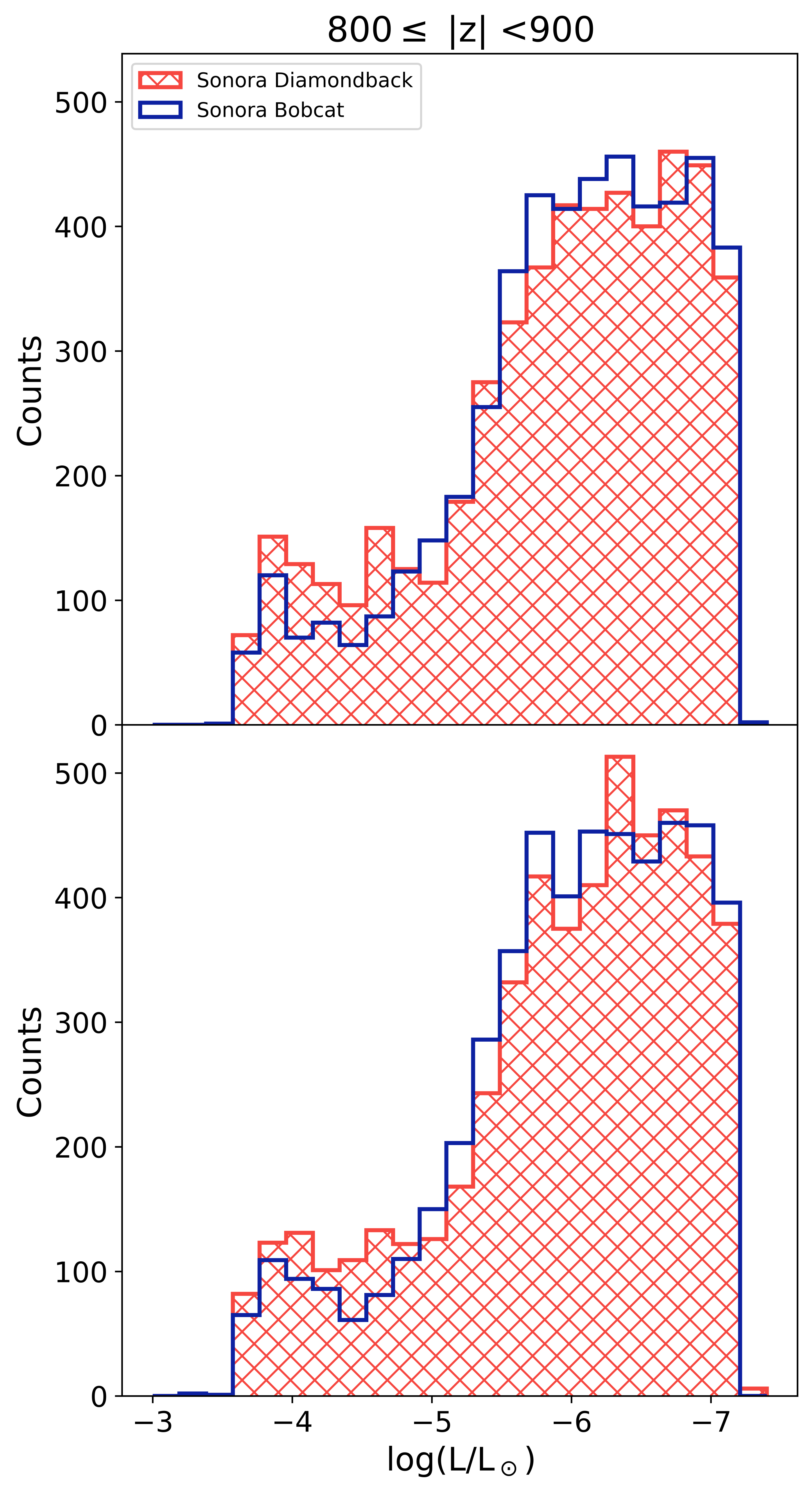}
\figsetgrpnote{Sonora luminosity functions for |z|=800-900 slice.}
\figsetgrpend

\figsetgrpstart
\figsetgrpnum{11.11}
\figsetgrptitle{$\abs{z}$ 900-1000 pc}
\figsetplot{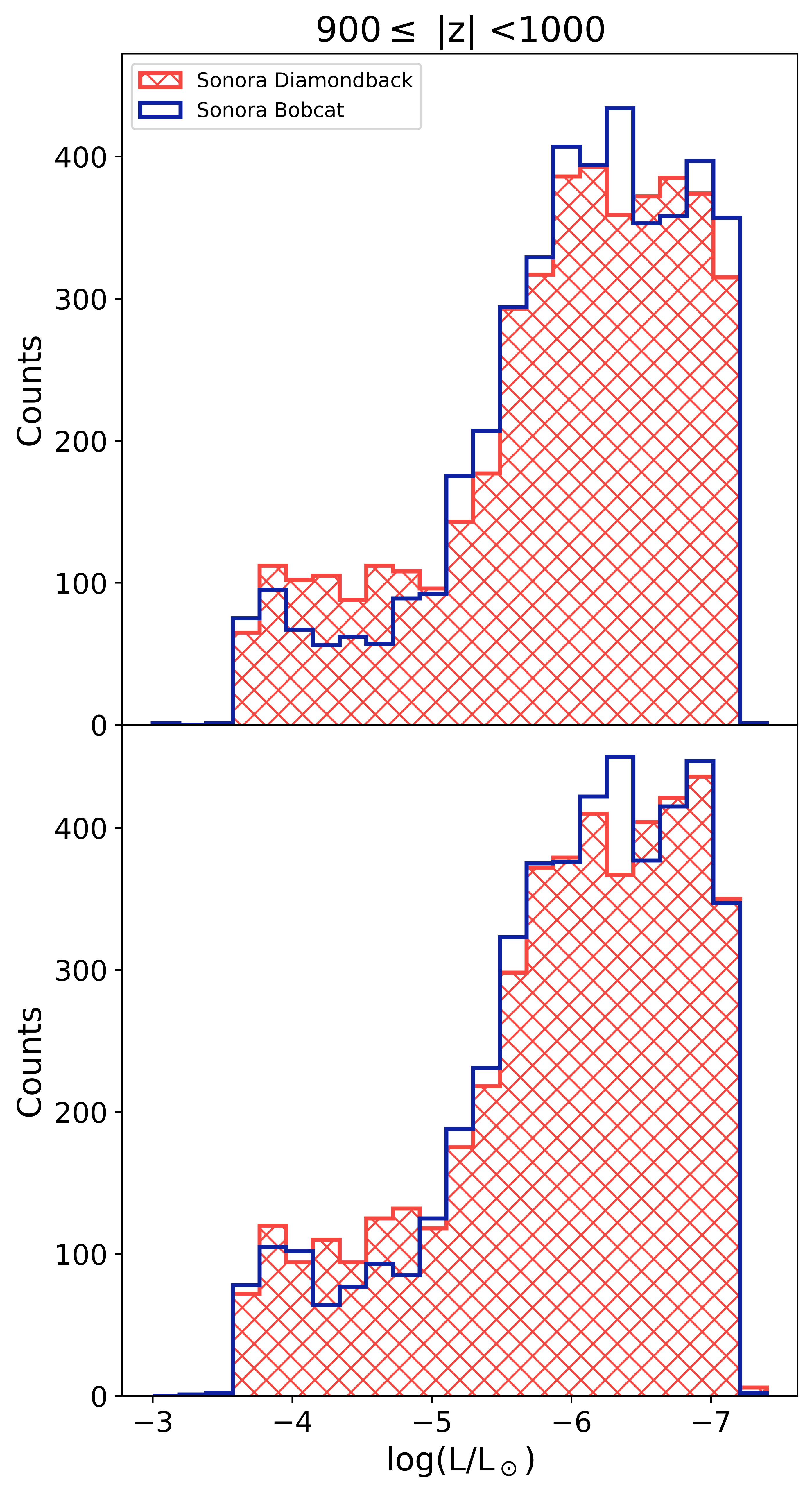}
\figsetgrpnote{Sonora luminosity functions for |z|=900-1000 slice.}
\figsetgrpend

\figsetgrpstart
\figsetgrpnum{11.12}
\figsetgrptitle{$\abs{z}$ 1000-1100 pc}
\figsetplot{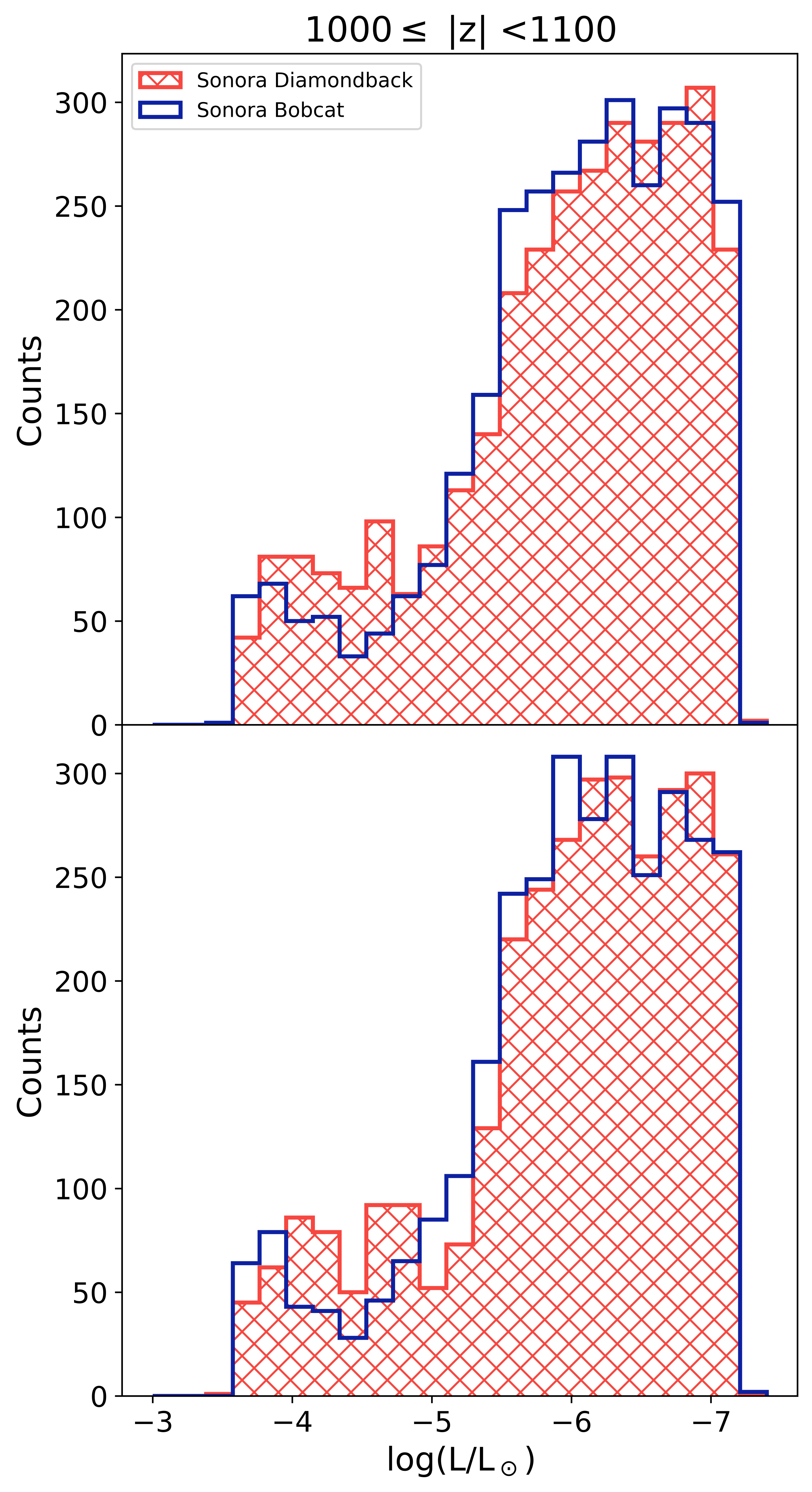}
\figsetgrpnote{Sonora luminosity functions for |z|=1000-1100 slice.}
\figsetgrpend

\figsetgrpstart
\figsetgrpnum{11.13}
\figsetgrptitle{$\abs{z}$ 1100-1200 pc}
\figsetplot{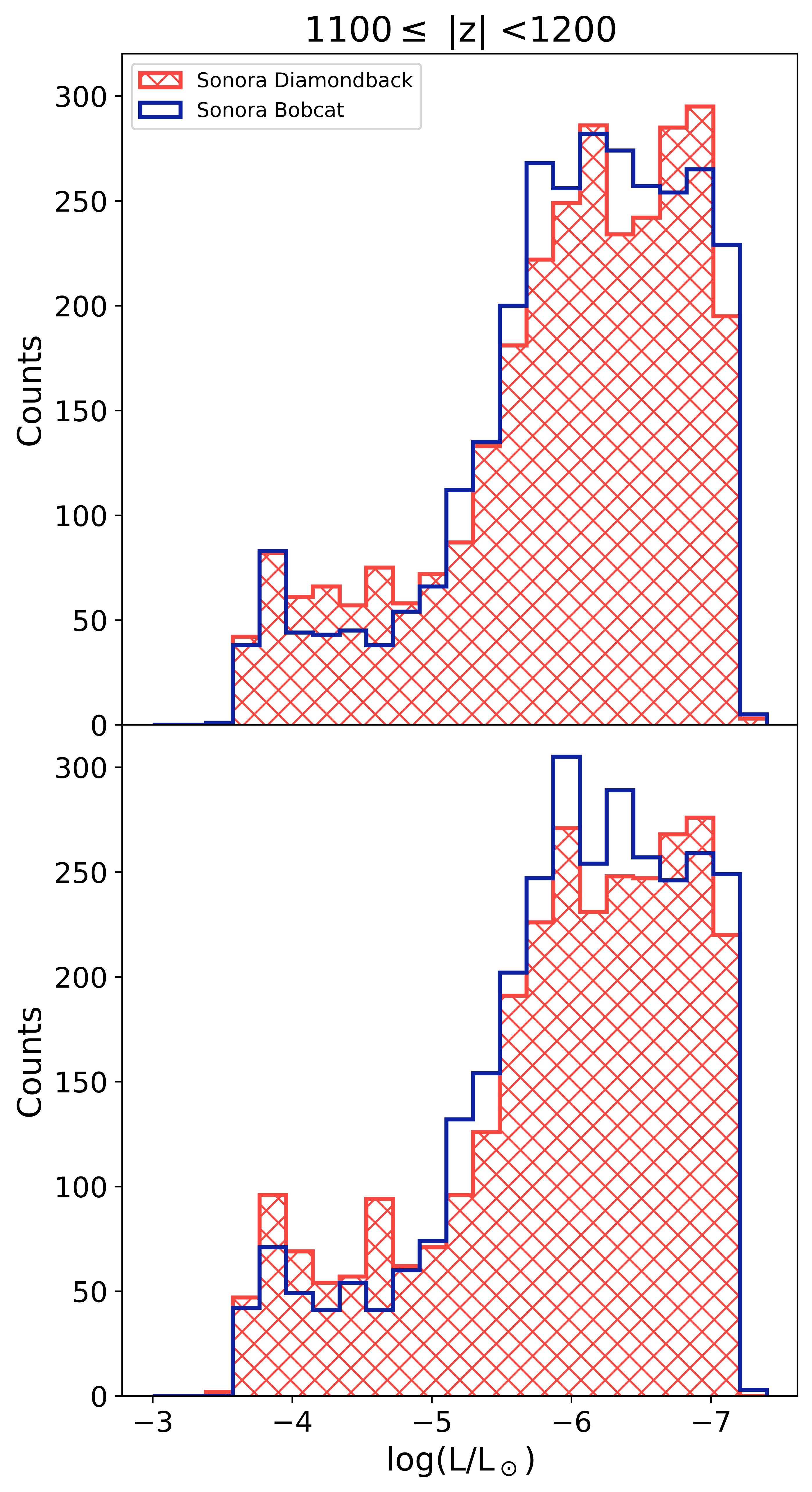}
\figsetgrpnote{Sonora luminosity functions for |z|=1100-1200 slice.}
\figsetgrpend

\figsetgrpstart
\figsetgrpnum{11.14}
\figsetgrptitle{$\abs{z}$ 1200-1300 pc}
\figsetplot{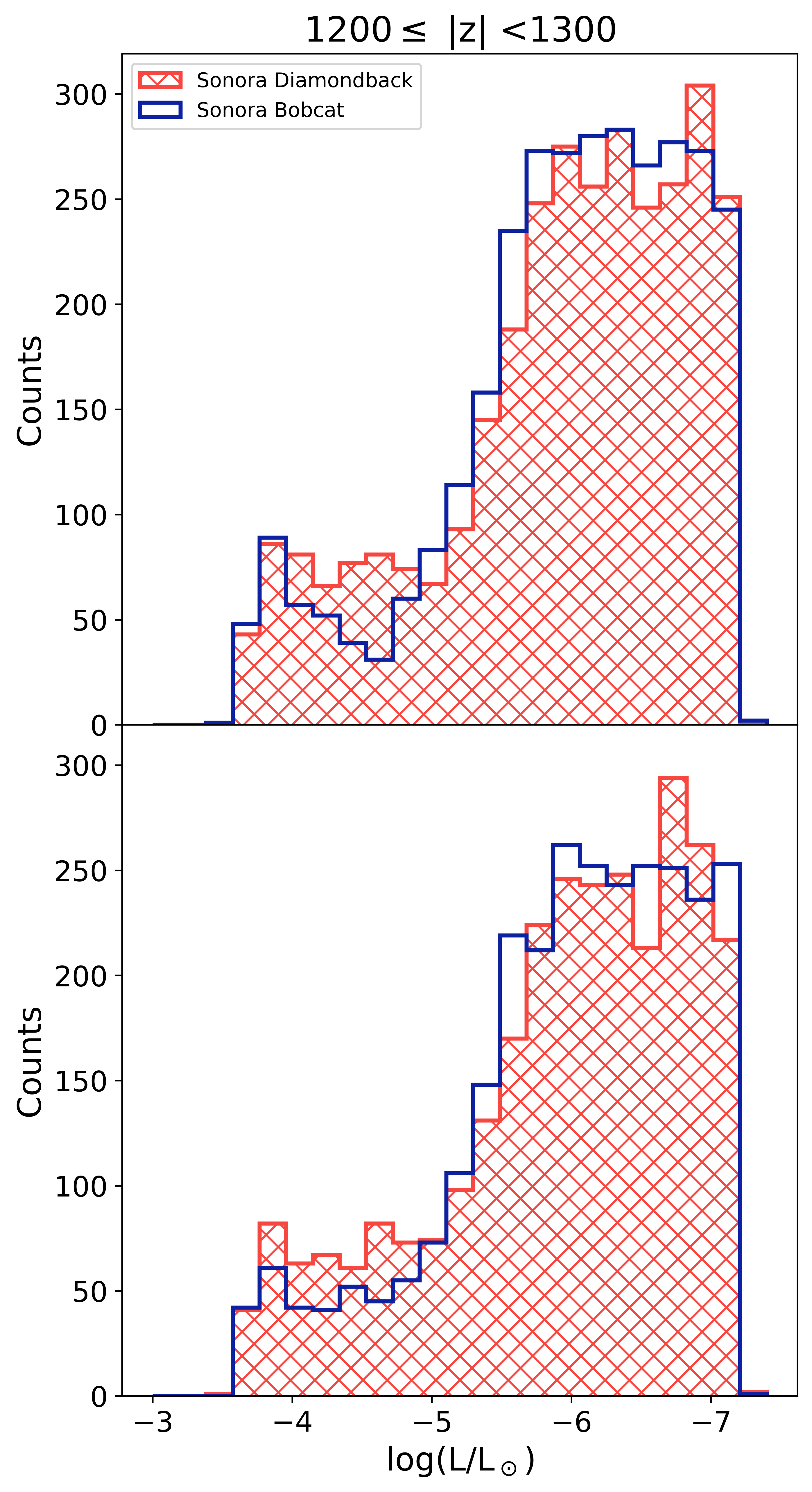}
\figsetgrpnote{Sonora luminosity functions for |z|=1200-1300 slice.}
\figsetgrpend

\figsetgrpstart
\figsetgrpnum{11.15}
\figsetgrptitle{$\abs{z}$ 1300-1400 pc}
\figsetplot{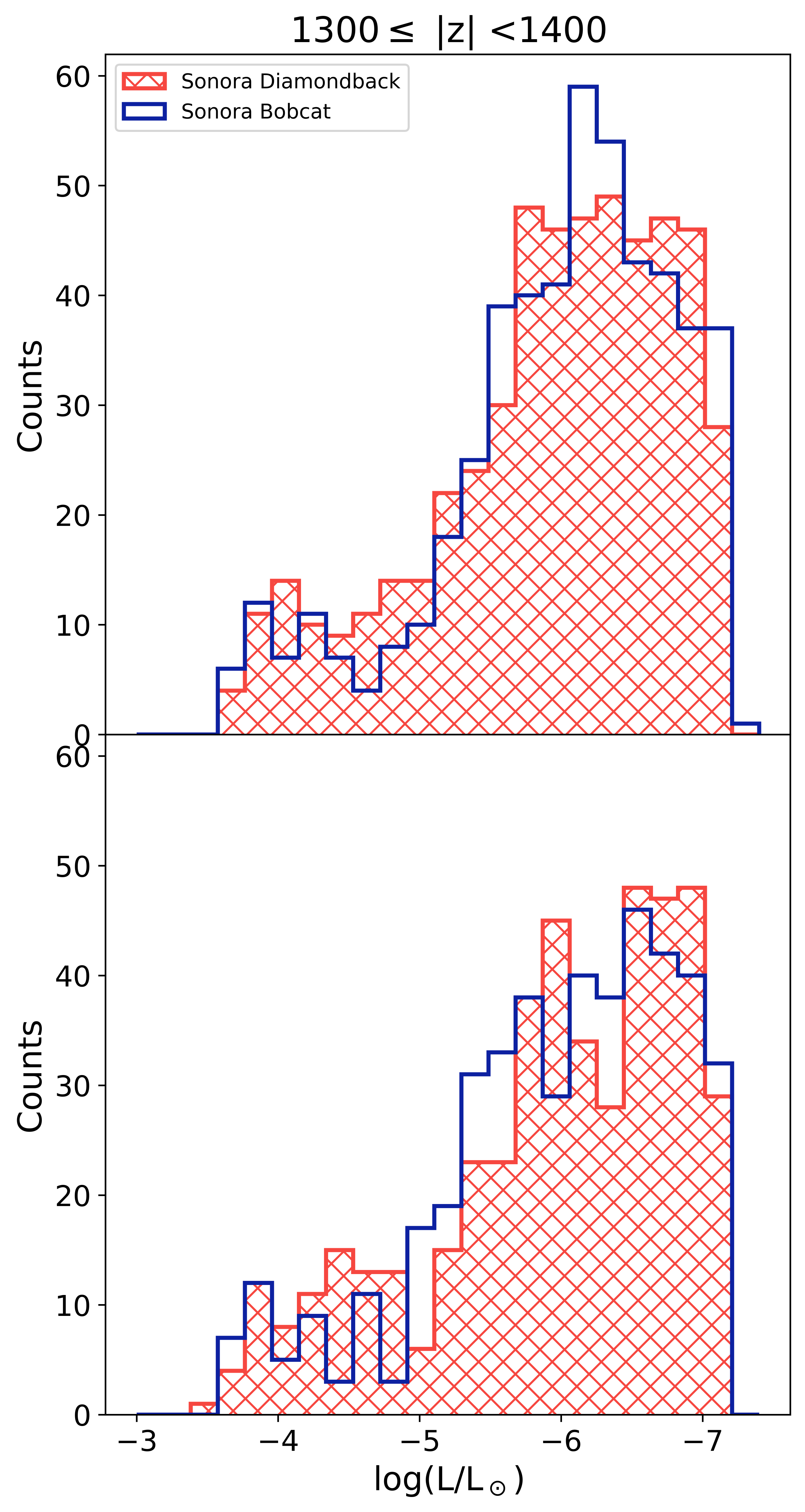}
\figsetgrpnote{Sonora luminosity functions for |z|=1300-1400 slice.}
\figsetgrpend

\figsetend

\begin{figure*}
    \centering
    \includegraphics[width = \textwidth]{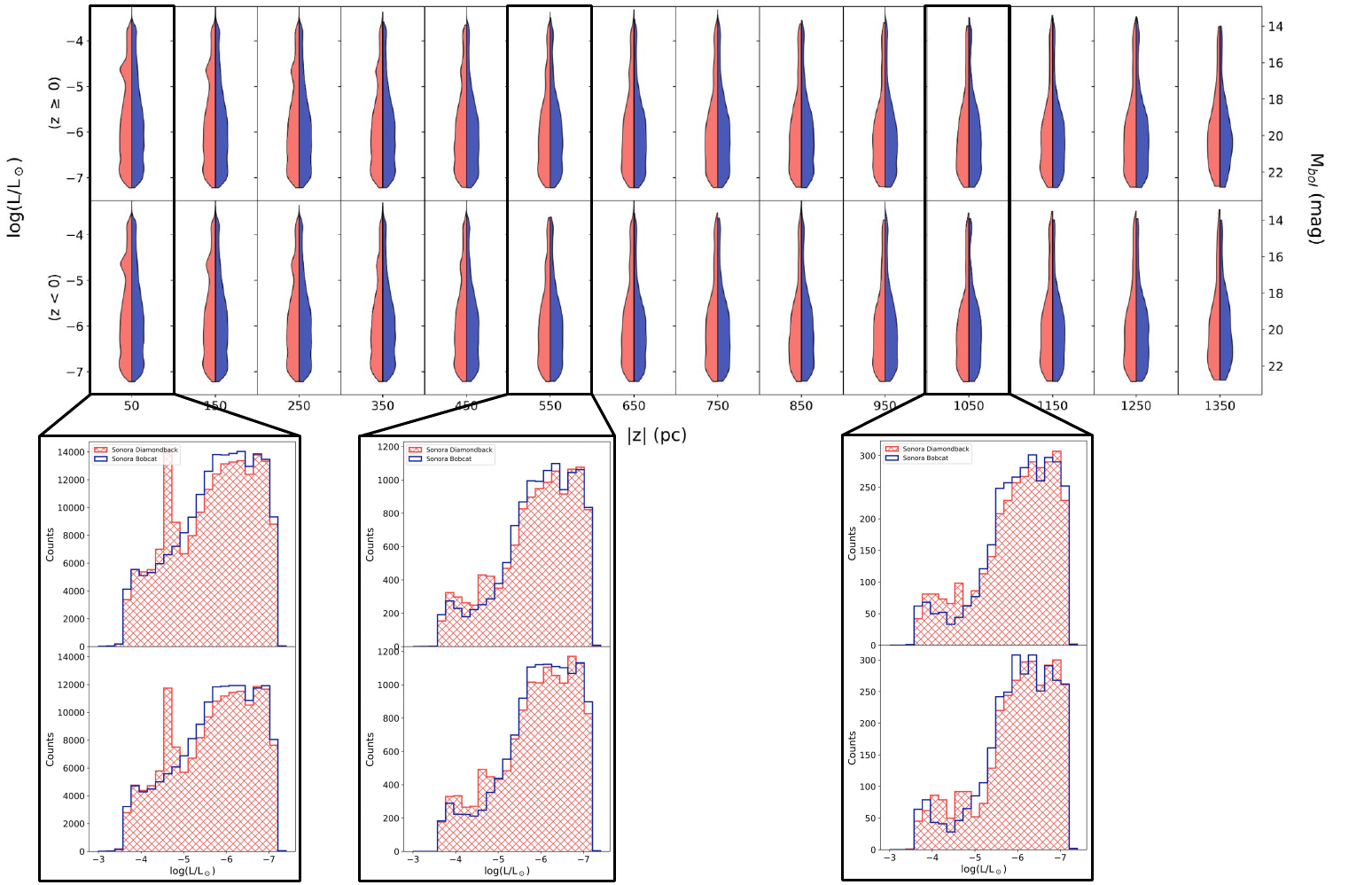}
    \caption{The luminosity function for the Sonora Diamondback (red) and Bobcat (blue) synthetic populations is shown in 100 pc bins. As distance from the Plane increases, the cloudy-cloudless transition feature in the Diamondback model dissipates and the population shifts to the lowest luminosities. In the slices closest to the Galactic Plane, the T-Y transition pileup is visible but is indistinguishable at larger distances. The Bobcat luminosity function is smooth but extends to brighter luminosities at lower $|z|$ slices. Three slices are shown in greater detail in pop-out histograms. The histograms closest to the Galactic Plane (far left) contain more objects and show the most prominent pileup features. The middle pop-out contains fewer objects and shows the decrease in pileup feature strength. The farthest pop-out from the Plane (far right) contains the least number of objects and the weakest pileup feature strength. The complete figure set, this figure and all slice pop-outs (15 figures), is available in the online journal.}
    \label{luminositywithpopouts}
\end{figure*}

Future surveys that will probe farther beyond the Galactic Plane will see combinations of these age distributions. Understanding the underlying relation between $\abs{z}$ and age is key to interpreting the observed luminosity function and evolution of the brown dwarf population. Volume-complete observational samples to date only extend to the nearest, youngest slices in our simulation. Local samples are biased against older objects, which is problematic in interpreting brown dwarf formation mechanisms. By extending our simulation to higher distances above/below the Galactic Plane, we include older objects that are underrepresented in the local sample. 

Coupled with age, our simulation includes metallicity predictions stemming from the \cite{daltio_AMR_2021} Age-Metallicity Relation (Equation \ref{AMR}). As a consequence of more distant slices being older on average, sub-populations farther from the Galactic Plane are more metal poor whereas the younger objects closer to the Plane are closer to Solar metallicity. Older populations are expected to be more metal poor than younger populations due to supernovae feedback, multiple stellar populations, and metallicity enrichment around the dusty, gaseous Galactic Plane. Our simulation is in good agreement with these expectations and we find the oldest sub-populations, most distant from the Plane, are more metal poor than the younger sub-populations near the Plane. However, our simulation can only probe the metallicity range from [M/H]$ =+0.5$ to $-0.5$ dex because of our choice for AMR. While this begins to explore the subdwarf domain (sd, $-1.0 \lesssim$[M/H]$\lesssim -0.3$ for L dwarfs, \citet{Zhang17}), we cannot probe the extreme subdwarf (esd, $-1.7\lesssim$[M/H]$\lesssim -1.0$, \citet{Zhang17}) or ultra subdwarf (usd, [M/H]$\lesssim -1.7$, \citet{Zhang17}) regimes. In order to accurately simulate the low-metallicity parameter space, a more robust AMR is necessary that can describe both the general brown dwarf population as well as the more rare, low metallicity objects.

In Figure \ref{luminositywithpopouts}, we show the evolution of the Sonora Diamondback and Bobcat luminosity functions in 100 pc intervals above/below the Galactic Plane. Similar to the spatially-resolved luminosity function for SM08 in Figure \ref{sm08_lum_heatmap}, the Sonora luminosity function morphologies and feature strengths change with $|z|$.  Specifically, the Diamondback cloudy-cloudless transition feature is prominent in sub-populations near the Galactic Plane but decreases in amplitude as $|z|$ increases, becoming nearly indistinguishable beyond 700 pc. Similarly, the T-Y transition is only visible within 400 pc of the Plane. For both Sonora models, the luminosity function morphology becomes more compact around lower luminosities as distance from the Plane increases.  

The detailed histograms of each 100 pc interval's luminosity function (shown as pop-outs in Figure \ref{luminositywithpopouts}), show the sub-population luminosity functions in finer detail and that the feature strengths and overall distribution changes significantly with $|z|$. Furthermore, intervals closest to the Galactic Plane contain the highest number of objects and are primarily younger. Thus, the contribution of slices near the Plane to the total luminosity function is greater and enhances the strength of the cloudy-cloudless transmission feature.

Since the luminosity function can be used to probe the cooling mechanisms and interior physics of brown dwarfs, improperly accounting for respective contributions from the more numerous, nearby young objects and the less numerous, distant older objects could lead to misinterpreting the physics of brown dwarfs. By including older populations more distant from the Galactic Plane, our simulation offers a novel approach to understanding a broader range of brown dwarf parameters than currently allowed by local samples. 

Understanding the evolution of the luminosity function with height above and below the Galactic Plane is of particular importance to accurately describe the ages, masses, and fundamental properties of ultracool dwarfs. The ages, masses, and luminosities of ultracool dwarfs are degenerate, so it is exceptionally challenging, if not impossible, to age these objects to understand their evolution \citep{Faherty14, dupuy17}. However, by dividing the Solar Neighborhood into respective slices with individual luminosity functions, the observed luminosity function can be compared with predicted functions and the contributions from different aged objects in various slices can be quantified.

\section{Summary}\label{SUMMARY}
We have simulated brown dwarfs in the Solar Neighborhood using Gaia-based star formation rate history both above and below the Galactic Plane. This novel approach allows us to measure the population statistics as a function of height and sample different age distributions. Normalized to the local brown dwarf spatial density, our simulation predicts the physical distribution, ages, and metallicities of brown dwarfs within a cylinder centered on the Galactic Plane. Our simulation predicts $\sim$750,000 brown dwarfs are encapsulated within such a cylinder and are concentrated near the Galactic Plane. We apply the hybrid SM08, gravity-dependent hybrid Sonora Diamondback, and cloudless Sonora Bobcat evolutionary models to our synthetic population to obtain effective temperatures, radii, surface gravities, and luminosities. The temperature and luminosity functions favor cooler, dimmer objects at all slices, but warmer, more luminous sub-populations appear near the Galactic Plane. The hybrid models include a characteristic pileup of objects around 1300 - 1400 K as the models transition from cloudy to cloudless. The pileup is a prominent feature for slices near the Galactic Plane but dissipates for more distant slices.

We explore the space densities and median ages for different spectral type ranges and find across all three synthetic populations that late-T dwarfs and Y dwarfs have the highest space densities. Similarly, Y dwarfs have the oldest median ages across the entire simulated volume, followed by late-T dwarfs. We compare the solar slice median ages for all three synthetic populations to published brown dwarf median ages and find our L dwarf and T dwarf median ages to be older than previously expected.

Understanding the underlying luminosity function as future surveys detect more distant objects is critical as the luminosity function is the only directly observable parameter -- other object parameters, such as temperature and radius, are model-derived quantities based on luminosity. By creating multiple evolutionary model luminosity functions, we can compare with the observed luminosity functions to understand brown dwarf evolution and cooling mechanisms. Our simulation accounts for older objects that are underrepresented in local samples, which adds significant, previously unavailable information for interpreting brown dwarf formation mechanisms. Furthermore, more distant populations within our simulation are older and more metal poor, a result that points to dynamical heating and Galactic dynamics.  

We reach a seemingly trivial, yet critical takeaway: \textit{As surveys resolve more distant populations of brown dwarfs, luminosity functions will sum along the line of sight. Therefore, it will become increasingly important to disentangle the contributions from older populations more distant from the Galactic Plane to properly understand brown dwarf cooling and formation mechanisms.} 

The next paper in this series will apply photometric survey footprints and nominal depths to the synthetic population to predict detection counts and population completeness. Future surveys such as JWST, Euclid, Rubin, and Roman will enable the statistical analysis of brown dwarf populations and will measure space densities as a function of height that can be compared to our predictions.

\section*{Acknowledgments}
The authors thank Leo Girardi and Alessandro Mazzi for email exchange in applying their results to this simulation. The authors also thank the anonymous referee for suggestions as well as the University of Delaware's Tali Natan, Sid Chaini, and FAST Lab for feedback on visualization approaches and figures. EJH is supported by NASA EPSCoR R3 award 80NSSC24M0160.

\textit{Software:}
\texttt{astropy} \citep{astropy2013,astropy2018,astropy2022}, \texttt{Jupyter} \citep{jupyter}, \texttt{NumPy} \citep{numpy},  \texttt{matplotlib} \citep{matplotlib}, \texttt{statsmodels} \citep{seabold}, \texttt{SciPy} \citep{Scipy}, and \texttt{Python3} \citep{python3}.

\appendix

In this Appendix, we present the median age and space density tables for the Sonora Bobcat and SM08 synthetic populations as mentioned in Section \ref{DISCUSSION}. The space densities and median ages for the Bobcat population are given in Tables \ref{bobcatspacedensities} \& \ref{bobcatmedianages}, respectively, while the space densities and median ages for the SM08 population are in Tables \ref{sm08spacedensities} \& \ref{sm08medianages}, respectively. 

\startlongtable
\begin{deluxetable*}{ c c|c c c c|c c c|c}
\tablecolumns{10}
\label{bobcatspacedensities}
\tablecaption{\centering{Space densities, $\rho$, by spectral type within \textit{z} slices for the Sonora Bobcat synthetic population. All space densities are reported in ($\times 10^{-3}$ pc$^{-3}$).}}
\tablehead{\colhead{z$_{\text{min}}$} & \colhead{z$_{\text{max}}$} & \colhead{L0-4} & \colhead{L5-9} & \colhead{T0-4} & \colhead{T5-9} & \colhead{L0-9} & \colhead{T0-9} & \colhead{Y0-2} & \colhead{Total}\\
\colhead{(pc)} & \colhead{(pc)} & \colhead{} & \colhead{} & \colhead{} & \colhead{} & \colhead{} & \colhead{} & \colhead{} & \colhead{} }
\startdata
-1315.78 & -1210.52 & 0.015 & 0.011 & 0.003 & 0.108 & 0.025 & 0.111 & 0.093 & 0.230 \\
-1210.52 & -1105.26 & 0.015 & 0.011 & 0.003 & 0.110 & 0.026 & 0.113 & 0.096 & 0.235 \\
-1105.26 & -1000.00 & 0.017 & 0.010 & 0.004 & 0.121 & 0.027 & 0.125 & 0.103 & 0.254 \\
-1000.00 & -894.74 & 0.026 & 0.018 & 0.005 & 0.172 & 0.044 & 0.177 & 0.152 & 0.372 \\
-894.74 & -789.47 & 0.027 & 0.017 & 0.007 & 0.195 & 0.045 & 0.201 & 0.164 & 0.410 \\
-789.47 & -684.21 & 0.033 & 0.020 & 0.009 & 0.220 & 0.053 & 0.229 & 0.195 & 0.477 \\
-684.21 & -578.95 & 0.048 & 0.033 & 0.013 & 0.322 & 0.081 & 0.335 & 0.270 & 0.685 \\
-578.95 & -473.68 & 0.075 & 0.059 & 0.022 & 0.541 & 0.134 & 0.563 & 0.441 & 1.138 \\
-473.68 & -368.42 & 0.120 & 0.106 & 0.033 & 0.806 & 0.225 & 0.839 & 0.639 & 1.704 \\
-368.42 & -263.16 & 0.165 & 0.152 & 0.052 & 1.111 & 0.317 & 1.163 & 0.876 & 2.356 \\
-263.16 & -157.89 & 0.286 & 0.254 & 0.084 & 1.737 & 0.540 & 1.822 & 1.364 & 3.725 \\
-157.89 & -105.26 & 0.506 & 0.476 & 0.157 & 2.775 & 0.983 & 2.932 & 2.128 & 6.042 \\
-105.26 & -52.63 & 0.857 & 0.826 & 0.260 & 4.443 & 1.683 & 4.703 & 3.274 & 9.660 \\
-52.63 & 0.00 & 1.424 & 1.362 & 0.413 & 6.926 & 2.787 & 7.339 & 5.176 & 15.302 \\
\hline
0.00 & 52.63 & 1.750 & 1.648 & 0.502 & 8.257 & 3.399 & 8.759 & 6.142 & 18.300 \\
\hline
52.63 & 105.26 & 0.985 & 0.935 & 0.310 & 5.004 & 1.920 & 5.314 & 3.733 & 10.967 \\
105.26 & 157.89 & 0.561 & 0.537 & 0.166 & 3.064 & 1.097 & 3.230 & 2.341 & 6.668 \\
157.89 & 263.16 & 0.314 & 0.285 & 0.098 & 1.783 & 0.599 & 1.881 & 1.393 & 3.873 \\
263.16 & 368.42 & 0.169 & 0.150 & 0.048 & 1.081 & 0.319 & 1.130 & 0.855 & 2.304 \\
368.42 & 473.68 & 0.100 & 0.080 & 0.029 & 0.664 & 0.180 & 0.693 & 0.535 & 1.407 \\
473.68 & 578.95 & 0.069 & 0.055 & 0.017 & 0.478 & 0.124 & 0.495 & 0.393 & 1.011 \\
578.95 & 684.21 & 0.055 & 0.041 & 0.014 & 0.369 & 0.096 & 0.383 & 0.311 & 0.791 \\
684.21 & 789.47 & 0.038 & 0.028 & 0.009 & 0.263 & 0.067 & 0.273 & 0.207 & 0.547 \\
789.47 & 894.74 & 0.024 & 0.019 & 0.007 & 0.189 & 0.043 & 0.196 & 0.162 & 0.401 \\
894.74 & 1000.00 & 0.022 & 0.015 & 0.004 & 0.161 & 0.037 & 0.165 & 0.140 & 0.342 \\
1000.00 & 1105.26 & 0.017 & 0.010 & 0.004 & 0.118 & 0.027 & 0.122 & 0.104 & 0.253 \\
1105.26 & 1210.52 & 0.016 & 0.010 & 0.003 & 0.109 & 0.026 & 0.113 & 0.095 & 0.233 \\
1210.52 & 1315.78 & 0.018 & 0.009 & 0.003 & 0.122 & 0.027 & 0.126 & 0.100 & 0.253 \\
\enddata
\end{deluxetable*}

\startlongtable
\begin{deluxetable*}{ c c | c c c c | c c c }
\tablecolumns{9}
\label{bobcatmedianages}
\tablecaption{Median ages by spectral type within \textit{z} slices for the Sonora Bobcat population.}
\tablehead{\colhead{z$_{\text{min}}$} & \colhead{z$_{\text{max}}$} & \colhead{L0-4} & \colhead{L5-9} & \colhead{T0-4} & \colhead{T5-9} & \colhead{L0-9} & \colhead{T0-9} & \colhead{Y0-2} \\
\colhead{(pc)} & \colhead{(pc)} & \colhead{(Gyr)} & \colhead{(Gyr)} & \colhead{(Gyr)} & \colhead{(Gyr)} & \colhead{(Gyr)} & \colhead{(Gyr)} & \colhead{(Gyr)} }
\startdata
-1315.78 & -1210.52 & 9.28 & 8.51 & 8.14 & 9.29 & 8.81 & 9.27 & 9.46 \\
-1210.52 & -1105.26 & 9.51 & 8.82 & 8.49 & 9.33 & 9.09 & 9.32 & 9.62 \\
-1105.26 & -1000.00 & 9.74 & 8.95 & 8.56 & 9.43 & 9.46 & 9.38 & 9.68 \\
-1000.00 & -894.74 & 8.91 & 8.54 & 7.51 & 8.97 & 8.79 & 8.95 & 9.36 \\
-894.74 & -789.47 & 8.90 & 8.09 & 7.17 & 9.00 & 8.58 & 8.96 & 9.36 \\
-789.47 & -684.21 & 8.71 & 8.21 & 6.85 & 8.94 & 8.50 & 8.89 & 9.22 \\
-684.21 & -578.95 & 8.14 & 6.58 & 6.44 & 8.61 & 7.54 & 8.55 & 8.88 \\
-578.95 & -473.68 & 7.81 & 5.55 & 5.65 & 8.16 & 6.84 & 8.09 & 8.53 \\
-473.68 & -368.42 & 6.39 & 4.45 & 4.39 & 7.24 & 5.39 & 7.16 & 8.01 \\
-368.42 & -263.16 & 5.99 & 3.99 & 4.22 & 6.99 & 5.03 & 6.87 & 7.83 \\
-263.16 & -157.89 & 5.09 & 3.14 & 3.53 & 6.38 & 3.98 & 6.25 & 7.24 \\
-157.89 & -105.26 & 3.31 & 2.65 & 2.97 & 5.66 & 2.93 & 5.48 & 6.68 \\
-105.26 & -52.63 & 2.86 & 2.35 & 2.61 & 5.11 & 2.56 & 4.86 & 6.29 \\
-52.63 & 0.00 & 2.50 & 2.17 & 2.53 & 5.08 & 2.31 & 4.82 & 6.41 \\
\hline
0.00 & 52.63 & 2.35 & 2.11 & 2.53 & 4.71 & 2.22 & 4.52 & 5.99 \\
\hline
52.63 & 105.26 & 2.68 & 2.29 & 2.67 & 5.37 & 2.45 & 5.12 & 6.71 \\
105.26 & 157.89 & 3.30 & 2.62 & 2.92 & 5.79 & 2.86 & 5.63 & 6.64 \\
157.89 & 263.16 & 4.44 & 2.85 & 3.14 & 5.97 & 3.25 & 5.82 & 6.88 \\
263.16 & 368.42 & 5.96 & 3.83 & 4.29 & 6.82 & 4.85 & 6.73 & 7.78 \\
368.42 & 473.68 & 7.19 & 5.13 & 4.90 & 7.65 & 6.29 & 7.52 & 8.28 \\
473.68 & 578.95 & 7.82 & 6.36 & 5.54 & 8.39 & 7.29 & 8.33 & 8.78 \\
578.95 & 684.21 & 7.94 & 6.41 & 5.24 & 8.25 & 7.34 & 8.18 & 8.76 \\
684.21 & 789.47 & 8.24 & 6.92 & 6.74 & 8.51 & 7.88 & 8.47 & 8.96 \\
789.47 & 894.74 & 8.79 & 7.28 & 6.49 & 8.73 & 8.22 & 8.70 & 9.19 \\
894.74 & 1000.00 & 9.28 & 8.52 & 8.55 & 9.14 & 9.11 & 9.12 & 9.52 \\
1000.00 & 1105.26 & 8.72 & 7.60 & 8.01 & 9.01 & 8.30 & 8.98 & 9.41 \\
1105.26 & 1210.52 & 8.99 & 8.90 & 8.37 & 9.33 & 8.95 & 9.31 & 9.67 \\
1210.52 & 1315.78 & 9.31 & 9.01 & 8.97 & 9.46 & 9.23 & 9.45 & 9.71 \\
\enddata
\end{deluxetable*}

\startlongtable
\begin{deluxetable*}{ c c|c c c c|c c c|c}
\tablecolumns{10}
\label{sm08spacedensities}
\tablecaption{\centering{Space densities, $\rho$, by spectral type within \textit{z} slices for the SM08 synthetic population. All space densities are reported in ($\times 10^{-3}$ pc$^{-3}$).}}
\tablehead{\colhead{z$_{\text{min}}$} & \colhead{z$_{\text{max}}$} & \colhead{L0-4} & \colhead{L5-9} & \colhead{T0-4} & \colhead{T5-9} & \colhead{L0-9} & \colhead{T0-9} & \colhead{Y0-2} & \colhead{Total}\\
\colhead{(pc)} & \colhead{(pc)} & \colhead{} & \colhead{} & \colhead{} & \colhead{} & \colhead{} & \colhead{} & \colhead{} & \colhead{} }
\startdata
-1315.78 & -1210.52 & 0.019 & 0.016 & 0.005 & 0.101 & 0.035 & 0.105 & 0.089 & 0.230 \\
-1210.52 & -1105.26 & 0.020 & 0.018 & 0.004 & 0.103 & 0.038 & 0.107 & 0.090 & 0.235 \\
-1105.26 & -1000.00 & 0.020 & 0.019 & 0.004 & 0.112 & 0.039 & 0.116 & 0.099 & 0.254 \\
-1000.00 & -894.74 & 0.030 & 0.028 & 0.008 & 0.162 & 0.059 & 0.170 & 0.144 & 0.372 \\
-894.74 & -789.47 & 0.030 & 0.033 & 0.009 & 0.180 & 0.063 & 0.188 & 0.158 & 0.410 \\
-789.47 & -684.21 & 0.040 & 0.039 & 0.011 & 0.208 & 0.079 & 0.220 & 0.178 & 0.477 \\
-684.21 & -578.95 & 0.054 & 0.059 & 0.018 & 0.301 & 0.113 & 0.319 & 0.254 & 0.685 \\
-578.95 & -473.68 & 0.088 & 0.105 & 0.032 & 0.498 & 0.193 & 0.530 & 0.415 & 1.138 \\
-473.68 & -368.42 & 0.136 & 0.166 & 0.051 & 0.735 & 0.302 & 0.786 & 0.615 & 1.704 \\
-368.42 & -263.16 & 0.195 & 0.235 & 0.072 & 1.004 & 0.430 & 1.075 & 0.850 & 2.356 \\
-263.16 & -157.89 & 0.308 & 0.395 & 0.123 & 1.588 & 0.703 & 1.711 & 1.311 & 3.725 \\
-157.89 & -105.26 & 0.498 & 0.731 & 0.211 & 2.558 & 1.229 & 2.769 & 2.045 & 6.042 \\
-105.26 & -52.63 & 0.835 & 1.233 & 0.365 & 3.976 & 2.068 & 4.341 & 3.251 & 9.660 \\
-52.63 & 0.00 & 1.399 & 2.019 & 0.579 & 6.203 & 3.418 & 6.782 & 5.102 & 15.302 \\
\hline
0.00 & 52.63 & 1.639 & 2.506 & 0.699 & 7.455 & 4.145 & 8.154 & 6.001 & 18.300 \\
\hline
52.63 & 105.26 & 0.979 & 1.39 & 0.400 & 4.542 & 2.369 & 4.942 & 3.656 & 10.967 \\
105.26 & 157.89 & 0.573 & 0.791 & 0.243 & 2.783 & 1.364 & 3.026 & 2.278 & 6.668 \\
157.89 & 263.16 & 0.325 & 0.435 & 0.133 & 1.635 & 0.761 & 1.768 & 1.345 & 3.873 \\
263.16 & 368.42 & 0.192 & 0.234 & 0.071 & 0.991 & 0.427 & 1.062 & 0.815 & 2.304 \\
368.42 & 473.68 & 0.114 & 0.131 & 0.042 & 0.611 & 0.245 & 0.653 & 0.509 & 1.407 \\
473.68 & 578.95 & 0.084 & 0.087 & 0.024 & 0.440 & 0.171 & 0.463 & 0.377 & 1.011 \\
578.95 & 684.21 & 0.065 & 0.061 & 0.023 & 0.347 & 0.126 & 0.370 & 0.294 & 0.791 \\
684.21 & 789.47 & 0.045 & 0.045 & 0.014 & 0.239 & 0.090 & 0.252 & 0.205 & 0.547 \\
789.47 & 894.74 & 0.035 & 0.030 & 0.007 & 0.179 & 0.065 & 0.187 & 0.150 & 0.401 \\
894.74 & 1000.00 & 0.030 & 0.025 & 0.005 & 0.150 & 0.055 & 0.155 & 0.133 & 0.342 \\
1000.00 & 1105.26 & 0.019 & 0.019 & 0.005 & 0.113 & 0.038 & 0.118 & 0.097 & 0.253 \\
1105.26 & 1210.52 & 0.019 & 0.017 & 0.005 & 0.103 & 0.036 & 0.108 & 0.090 & 0.233 \\
1210.52 & 1315.78 & 0.021 & 0.019 & 0.006 & 0.106 & 0.040 & 0.112 & 0.101 & 0.253 \\
\enddata
\end{deluxetable*}

\startlongtable
\begin{deluxetable*}{ c c | c c c c | c c c }
\tablecolumns{9}
\label{sm08medianages}
\tablecaption{\centering{Median ages by spectral type within \textit{z} slices for the SM08 population.}}
\tablehead{\colhead{z$_{\text{min}}$} & \colhead{z$_{\text{max}}$} & \colhead{L0-4} & \colhead{L5-9} & \colhead{T0-4} & \colhead{T5-9} & \colhead{L0-9} & \colhead{T0-9} & \colhead{Y0-2} \\
\colhead{(pc)} & \colhead{(pc)} & \colhead{(Gyr)} & \colhead{(Gyr)} & \colhead{(Gyr)} & \colhead{(Gyr)} & \colhead{(Gyr)} & \colhead{(Gyr)} & \colhead{(Gyr)} }
\startdata
-1315.78 & -1210.52 & 9.12 & 8.32 & 8.21 & 9.33 & 8.88 & 9.30 & 9.76 \\
-1210.52 & -1105.26 & 9.40 & 8.87 & 8.47 & 9.40 & 9.17 & 9.35 & 9.64 \\
-1105.26 & -1000.00 & 9.47 & 9.08 & 8.34 & 9.39 & 9.31 & 9.36 & 9.69 \\
-1000.00 & -894.74 & 9.13 & 8.38 & 6.72 & 9.06 & 8.64 & 9.00 & 9.39 \\
-894.74 & -789.47 & 9.25 & 8.20 & 6.55 & 9.12 & 8.71 & 9.06 & 9.37 \\
-789.47 & -684.21 & 9.32 & 8.06 & 7.55 & 8.85 & 8.67 & 8.78 & 9.31 \\
-684.21 & -578.95 & 8.75 & 6.83 & 5.90 & 8.62 & 8.15 & 8.53 & 8.97 \\
-578.95 & -473.68 & 8.11 & 5.73 & 5.07 & 8.16 & 7.04 & 8.04 & 8.56 \\
-473.68 & -368.42 & 7.70 & 5.17 & 4.16 & 7.43 & 6.16 & 7.21 & 7.87 \\
-368.42 & -263.16 & 7.14 & 4.42 & 4.20 & 7.18 & 5.65 & 6.94 & 7.69 \\
-263.16 & -157.89 & 6.45 & 3.31 & 3.26 & 6.55 & 4.62 & 6.29 & 7.19 \\
-157.89 & -105.26 & 5.26 & 2.73 & 2.92 & 5.82 & 3.12 & 5.55 & 6.34 \\
-105.26 & -52.63 & 4.09 & 2.46 & 2.79 & 5.23 & 2.76 & 4.90 & 5.87 \\
-52.63 & 0.00 & 3.31 & 2.25 & 2.53 & 5.34 & 2.54 & 4.95 & 5.98 \\
\hline
0.00 & 52.63 & 3.11 & 2.18 & 2.56 & 4.88 & 2.45 & 4.54 & 5.53 \\
\hline
52.63 & 105.26 & 4.02 & 2.41 & 2.78 & 5.54 & 2.74 & 5.21 & 6.24 \\
105.26 & 157.89 & 5.11 & 2.65 & 2.88 & 5.90 & 3.02 & 5.62 & 6.51 \\
157.89 & 263.16 & 5.88 & 2.97 & 3.15 & 6.10 & 3.84 & 5.83 & 6.69 \\
263.16 & 368.42 & 7.21 & 4.21 & 4.13 & 6.93 & 5.56 & 6.70 & 7.69 \\
368.42 & 473.68 & 7.92 & 5.20 & 4.87 & 7.68 & 6.39 & 7.47 & 8.34 \\
473.68 & 578.95 & 8.59 & 6.41 & 5.77 & 8.36 & 7.87 & 8.28 & 8.84 \\
578.95 & 684.21 & 8.74 & 6.51 & 5.45 & 8.38 & 8.05 & 8.25 & 8.81 \\
684.21 & 789.47 & 8.56 & 7.37 & 6.46 & 8.50 & 8.10 & 8.46 & 8.86 \\
789.47 & 894.74 & 9.20 & 7.80 & 5.96 & 8.79 & 8.78 & 8.73 & 9.06 \\
894.74 & 1000.00 & 9.49 & 8.48 & 7.90 & 9.19 & 9.07 & 9.17 & 9.59 \\
1000.00 & 1105.26 & 8.71 & 8.20 & 8.00 & 9.07 & 8.54 & 9.02 & 9.32 \\
1105.26 & 1210.52 & 9.37 & 8.51 & 8.33 & 9.38 & 8.90 & 9.33 & 9.64 \\
1210.52 & 1315.78 & 9.24 & 9.23 & 8.47 & 9.32 & 9.23 & 9.32 & 9.59 \\
\enddata
\end{deluxetable*}

\bibliography{main.bib}{}
\bibliographystyle{aasjournal}
\end{document}